\documentclass[journal]{IEEEtran}
\usepackage{graphicx}
\usepackage{cite}
\usepackage{amsmath,amssymb,amsfonts}
\usepackage{algorithmic}
\usepackage{graphicx}
\usepackage{textcomp}
\usepackage{mathrsfs}
\usepackage{epsfig}
\usepackage{mathptmx}
\usepackage{times}
\usepackage{mathtools}
\usepackage{booktabs}
\usepackage{amsmath}
\usepackage{float} 
\usepackage{dutchcal}
\usepackage{amssymb}
\usepackage{color}
\usepackage{subfigure}

\begin{document}
\title{Toward Reliable Designs of Data-Driven Reinforcement Learning Tracking Control for Euler-Lagrange Systems}

\author{	
	Zhikai Yao,
	Jennie Si,~\IEEEmembership{Fellow,~IEEE}, 
	Ruofan Wu,
	and Jianyong Yao,~\IEEEmembership{Member,~IEEE}
	\thanks{This work is supported in part by National Science Foundation \#1563921, and \#1808752. \textit{(Corresponding authors: Jennie Si and Jianyong Yao.)}}
	\thanks{Z. Yao is with the School of Mechanical Engineering, Nanjing University of Science and Technology, Nanjing,
		Jiangsu Province, 210094 CN, and also with the Department of Electrical, Computer, and Energy Engineering, Arizona State University, Tempe, AZ, 85281, USA (e-mail: zacyao.cn@gmail.com).}
	\thanks{J. Si and R. Wu are with the Department of Electrical, Computer, and Energy Engineering, Arizona State University, Tempe, AZ, 85281, USA (e-mail: si@asu.edu; ruofanwu@asu.edu).}
	\thanks{J. Yao is with the School of Mechanical Engineering, Nanjing University of Science and Technology, Nanjing,
		Jiangsu Province, 210094 CN (e-mail: jerryyao.buaa@gmail.com).}
}

%\markboth{IEEE Transactions on Neural Networks and Learning Systems}%

\maketitle

\begin{abstract}
This paper addresses reinforcement learning based, direct signal tracking control with an objective of developing mathematically suitable and practically useful design approaches. Specifically, we aim to provide reliable and easy to implement designs in order to reach reproducible neural network-based solutions. Our proposed new design takes advantage of two control design frameworks: a reinforcement learning based, data-driven  approach to provide the needed adaptation and (sub)optimality, and a backstepping based approach to provide closed-loop system stability framework. We develop this work based on an established direct heuristic dynamic programming (dHDP) learning paradigm to perform online learning and adaptation and a backstepping design for a class of important nonlinear dynamics described as Euler-Lagrange systems. We provide a theoretical guarantee for the stability of the overall dynamic system, weight convergence of the approximating nonlinear neural networks, and the Bellman (sub)optimality of the resulted control policy. We use simulations to demonstrate significantly improved design performance of the proposed approach over the original dHDP.

\end{abstract}


\IEEEpeerreviewmaketitle

\section{Introduction}

\IEEEPARstart{W}{e} consider the problem of data-driven optimal tracking control for Euler-Lagrange systems which are represented in a wide range of application problems. The evolution of such systems can be described by the solutions to the Euler-Lagrange equation. In classical mechanics, it is equivalent to Newton's laws of motion. Such systems also have the advantage that they use the same generalized coordinate system that makes solving the solution of motion easier. The Euler-Lagrange mechanisms can be found in many familiar systems, such as marine navigation equipment, automatic machine tools, satellite-tracking antennas, remote control airplanes, automatic navigation systems on boats and planes, autofocus cameras, computer hard disk drive, and more. In the modern computer and control era, such mechanisms are still behind the robotic manipulators, wearable robots, ground vehicles, and many more applications in mechanical, electrical and electromechanical systems.

Tracking control has been studied extensively in control theoretic context where the tracking control designs are based on well defined mathematical models of the  nonlinear dynamics. Well-established approaches include backstepping control \cite{krstic1995nonlinear}, observer-based control \cite{khalil2002nonlinear} and nonlinear
adaptive/robust control \cite{nijmeijer1990nonlinear, isidori2013nonlinear}. Those important results have provided the foundation for nonlinear tracking control, yet, their applicability may be limited especially when the nonlinear dynamics are difficult or impossible to precisely model. As such, data-driven nonlinear tracking control designs are sought after. Common and natural approaches are the use of machine learning or neural networks techniques as they can learn directly from data by means of the universally approximating property.

Most of the existing data-driven tracking control results focus on stabilization of nonlinear dynamic systems \cite{sun2019novel}. As is well-known, real engineering applications require considerations of optimal control performance, not just stability to account for factors such as energy consumption, tracking error rate, and more. Therefore, optimal tracking control solutions of nonlinear systems are sought after. A classical formulation of the problem is to obtain nonlinear optimal tracking control solutions from solving the Hamilton-Jacobi-Bellman (HJB) equation. But solving the HJB equation poses great challenges for general nonlinear systems. One such challenge is a lack of closed-form analytical solution even if a mathematical description of the nonlinear dynamics is available. Additionally, traditional approaches to solving the HJB equations are backward in time and therefore, can only be solved offline. Data-driven nonlinear optimal tracking control designs provide new promises to address these challenges, yet, they face new obstacles.

Currently, there is only a handful of results concerning the theory, algorithm design and implementation of data-driven optimal tracking control. Central to these results  \cite{fan2018robust,fu2020mrac,radac2019data,zhang2008novel, yang2009direct, zhang2011data, wei2014adaptive, modares2014optimal, kiumarsi2015actor, kamalapurkar2015approximate, modares2015h, luo2016model, mu2017data, gao2018learning, wang2018neural, zhao2019event, mu2019learning, dong2020optimal, ni2013adaptive} are the use of reinforcement learning to establish an approximate solution to the HJB equation. Yet, few results have demonstrated that these methods are not only mathematically suitable but also practically useful in order to address real engineering problems in terms of providing reliable, easy to implement designs that lead to reproducible neural network-based solutions. In addressing the design of reinforcement learning based optimal tracking of coal gasification problem, the authors of \cite{wei2014adaptive} first applied offline  neural network identifications to establish the necessary mathematical descriptions of the nonlinear system dynamics and the desired tracking trajectory in order to carry on the control design. As such, it is questionable if approaches based on similar ideas can potentially be useful for other applications as obtaining reproducible models will be the first barrier to overcome. This is not a trivial problem, as it requires great expertise and the subject is still under investigation because current neural network modeling results usually introduce large variances which depend on the designer and the hyperparameters used in learning the models. The authors of \cite{ni2013adaptive} proposed a tracking control solution based on dHDP \cite{si2001online} by making use of an additional neural network to provide an internal goal signal. This is theoretically suitable but it complicates the problem as discussed previously. The approximation errors are on top of the uncertainties introduced as approximation variances due to dHDP tracking solutions as pointed out in \cite{yang2009performance}. As such, the reproducibility of the approach is yet to be demonstrated systematically.

Among those existing reinforcement learning based tracking control design, they often rely on a reference model from which a (continuously differential) desired tracking signal $x_{\rm{d}}$ can be obtained \cite{fan2018robust,fu2020mrac,radac2019data,zhang2008novel, yang2009direct, zhang2011data, wei2014adaptive, modares2014optimal, kiumarsi2015actor, kamalapurkar2015approximate, modares2015h, luo2016model, mu2017data, gao2018learning, wang2018neural, zhao2019event, mu2019learning, dong2020optimal}. While it may be feasible and useful for certain applications such as flight validation \cite{Nhan2018Model}, the issue of constructing an appropriate reference model for general nonlinear system control purposes has not been thoroughly addressed. Actually, few published results are available either from a general theoretic perspective or from specific applications perspective. Even for a specific application, it is not considered an easy task \cite{da2019choice}. It is therefore fair to say that choosing an appropriate reference model is quite challenging and it may have been taken for granted. This could be the reason that most results on tracking related work have focused on addressing tracking control algorithm design or improving convergence properties of tracking algorithms. As one can imagine, the problem can exacerbate for large scale, complex dynamic systems. Even worse for some applications, it is nearly impossible to accurately capture nonlinear dynamics using a mathematical model. It is also worth mentioning that, some reported results \cite{fan2018robust,fu2020mrac,radac2019data,zhang2008novel, yang2009direct, zhang2011data, wei2014adaptive, modares2014optimal, kiumarsi2015actor, kamalapurkar2015approximate, modares2015h, luo2016model, mu2017data, gao2018learning, wang2018neural, zhao2019event, mu2019learning, dong2020optimal} require generating a corresponding reference control, which can also be challenging and make some of these approaches less applicable. Another commonly implied assumption in many existing learning based tracking control is that the nonlinear dynamics are partially known (or specifically the input dynamics are known). This has simplified the problem but in the meantime, it limits their applicability.

To directly use a reference trajectory in place of a model reference structure, backstepping idea can be employed as it allows for the construction of both feedback control laws and associated Lyapunov functions in a systematic way \cite{krstic1995nonlinear}. The backstepping idea in this context was examined in \cite{zargarzadeh2015optimal, wang2016backstepping} with critic-only reinforcement learning control. However, their results are limited to partially known system dynamics. Alternatively, the dHDP construct allows for direct use of reference trajectory as well in the design of tracking control. The idea was demonstrated via simulations in \cite{yang2009direct, ni2013adaptive}. The results are promising, yet, they lack a theoretical support for (sub)optimality and stability analysis. Even though tracking control results in \cite{yang2009direct} were obtained for dHDP as well, our current approach is fundamentally different as we propose a new control strategy and we use an informative tracking error based stage cost instead of a binary cost. Specifically, the current study is motivated to circumvent some long-standing issues in reinforcement learning, including $Q$-learning and dHDP, that is to reduce the variances in the resulted action and/or critic networks after training and thus to improve the reproductivity and consistency of the trained actor/critic network outcomes even if they are trained by different designers.

In this paper, we aim at developing a new nonlinear tracking control design framework with a goal of making the design approach feasible for applications. In a previous study \cite{yang2009performance}, we have shown that well initialized actor-critic neural networks in dHDP can significantly improve the quality of the optimal value function approximation and the optimal control policy. From the same study we realized the importance of finding an initially good estimate of a (locally) optimal solution to the Bellman equation. In this study, we create a backstepping control strategy to provide a feedback system stability framework and a dHDP online adaptation control strategy to provide a feed-forward compensation in order to obtain near optimal tracking control solution. In the feedback control performance objective, we take into account the feed-forward control input. As a result of this innovative solution approach, we can provide an overall system stability guarantee and also avoid the use of a reference model for the desired tracking trajectory. As such, we avoid the challenge of creating a reference model and also  remove one source of approximation error. Because of using a reinforcement learning feed-forward control, we address unknown nonlinear dynamics via learning from data, i.e., our design is data-driven, not model dependent.

Our contributions of this work are as follows.
\begin{itemize}
	\item [1)] 
We introduce a new reinforcement learning control design
framework within a backstepping feedback construct. This allows us to avoid the need of a fully identified system model for backstepping-based control as well as the dependence on a reference model for the desired tracking trajectory for reinforcement learning based-control. Additionally, the backstepping feedback component provides a guideline to narrow down the representation domain to be explored by neural networks in dHDP and to increase the chance of reaching a good (sub)optimal solution.
      
	\item [2)]
    We provide a theoretical guarantee for the stability of the overall dynamic system, weight convergence of the approximating nonlinear neural networks, and the Bellman (sub)optimality of the resulted control policy. 
	\item [3)]
	We provide simulations to not only demonstrate how the proposed design method works but also to show how the proposed algorithm can significantly improve reproducibility of the results under dHDP integrated with backstepping feedback stabilizing control.
\end{itemize}

The rest of the paper is organized as follows. Section \uppercase\expandafter{\romannumeral2} provides the problem formulation. Section \uppercase\expandafter{\romannumeral3} presents the backstepping design. Section \uppercase\expandafter{\romannumeral4} develops the reinforcement learning control. Section \uppercase\expandafter{\romannumeral5} provides theoretical analyses of the proposed algorithm. Simulation and comparison results are presented in Section \uppercase\expandafter{\romannumeral6} and the concluding remarks are given in Section \uppercase\expandafter{\romannumeral7}.

\section{Problem Formulation}
We consider a class of nonlinear dynamics described as Euler-Lagrange systems that govern the motion of rigid structures:
\begin{eqnarray}
\begin{aligned}
M(q(t)) \ddot{q}(t) +& V_{m}(q(t), \dot{q}(t)) \dot{q}(t) \\
+ &G(q(t)) + F(\dot{q}(t)) + \tau_{d}(t) = \tau(t).
\end{aligned}
\end{eqnarray}

In Eq. (1), $M(q(t))$, $V_{m}(q(t), \dot{q}(t))$, $G(q(t))$, $F(\dot{q}(t))$ and $\tau_{d}(t)$ are unknown, $q(t), \dot{q}(t), \ddot{q}(t) \in \mathbb{R}^{n}$ denote the rigid link position, velocity, and acceleration vectors, respectively; $M(q(t)) \in \mathbb{R}^{n \times n}$ denotes the inertia matrix; $V_{m}(q(t), \dot{q}(t)) \in \mathbb{R}^{n \times n}$ the centripetal-coriolis matrix; $G(q(t)) \in \mathrm{R}^{n}$ the gravity vector, $F(\dot{q}(t)) \in \mathbb{R}^{n}$ friction; $\tau_{d}(t) \in \mathbb{R}^{n}$ a disturbance; and $\tau(t) \in \mathbb{R}^{n}$ represents the torque control input. The subsequent development is based on the assumption that $q(t)$ and $\dot{q}(t)$ are measurable.

To carry on the development of tracking control of the dynamics in Eq. (1), we use the following general discrete-time state space representations where $x_1(k)\in \mathbb{R}^{n}$ and $x_2(k)\in \mathbb{R}^{n}$ denote link position and velocity, respectively. Applying the Euler discretization as in \cite{miranda2017multivalued}, systems as described by Eq. (1) can be rewritten as 
\begin{eqnarray}
\begin{aligned}
x_{1}(k+1) &= h x_2(k) + x_1(k) \\
M^{+}(k)x_{2}(k+1) &= u(k) - g(k) - \tau_{d}(k),
\end{aligned}
\end{eqnarray}
where $h = t_{k+1} - t_{k}$, $u(k) = \tau(t_k)$, $g(k)=V_{m}(k) x_{2}(k) + G(k) + F(k) - M^{-}(k)x_2(k)$, $M^{+}(k)$ denotes $M(t_{k+1})/h$ and $M^{-}(k)$ denotes $M(t_{k})/h$. The control objective is for the output $x_{1}(k)$ to track a desired time-varying trajectory $x_{1\rm{d}}(k)$ as closely as possible.

\textbf{Assumption 1.} The inertia matrix $M(q)$ is symmetric, positive definite, and the following inequality holds:
\begin{eqnarray}
 M_{min} \leq \| (M^{-}(k))^{T}(M^{-}(k)) \|,
\end{eqnarray}
where $M_{min}$ is a given positive constant, and $\|\cdot\|$ denotes the Euclidean norm.

\textbf{Assumption 2.} The nonlinear disturbance term $\tau_{d}(k)$ is bounded, i.e., $\tau_{d}(k) \in \mathcal{L}_{\infty}$.

In subsequent development, the desired trajectory $x_{1\rm{d}}(k)$ is not defined by any reference model, but rather, $x_{1\rm{d}}(k)$ is simply and directly provided. Our proposed new closed-loop, data-driven nonlinear tracking control solution is as shown in Fig. 1. There are two major components in the design. The backstepping design provides a feedback control structure, the feed-forward signal from the backstepping framework is then accounted for by using a dHDP online learning scheme. Even though the dHDP alone can theoretically be used for tracking control while preserving some qualitative properties such as those in \cite{yang2009direct}, our current approach aims at improving reliability of the design and reproducibility of the results while maintaining theoretical suitability. Next, we provide a comprehensive introduction of the two constituent control design blocks.     
\section{Backstepping To Provide Baseline Tracking Control}
The control goal is to find $u(k)$ (Fig.1) so that the output $x_1(k)$ of the system Eq. (2) tracks the desired time-varying trajectory $x_{1 \mathrm {d}}(k)$. The discrete-time backstepping scheme for the tracking problem of Eq. (2) with unknown, nonlinear system dynamics is derived step by step as follows. Refer to Fig.1, the two blocks play complementary roles in constructing the final control signal $u(k)$. The  backstepping control block is designed as follows.
%based on a rough estimated of the inertia $M^{+}(k)$ that satisfies (3). Specific design approaches of this block is

\textbf{Step 1.} To develop the backstepping design, a virtual control function is synthesized. Let $e_{1}(k)\in \mathbb{R}^{n}$ be the deviation of $x_1(k)$ from the target $x_{1 \mathrm {d}}(k) \in \mathbb{R}^{n}$, i.e., 
\begin{eqnarray}
e_{1}(k) = x_{1}(k) - x_{1 \mathrm {d}}(k). 
\end{eqnarray}

From (2) and (4), we have
\begin{eqnarray}
e_{1}(k+1) = h x_{2}(k) + x_{1}(k) - x_{1 \mathrm {d}}(k + 1).
\end{eqnarray}
Then, we view $x_{2}(k)$ as a virtual control in Eq. (5) and introduce the error variable 
\begin{eqnarray}
e_{2}(k) = x_{2}(k) - \alpha(k),
\end{eqnarray}
where $\alpha(k) \in \mathbb{R}^{n}$ is a stabilizing function for $x_{2}(k)$ to be chosen as
\begin{eqnarray}
\begin{aligned}
\alpha(k) = \frac{1}{h} (c_{1} e_{1}(k) - x_{1}(k) + x_{1 \mathrm {d}}(k+1)),
\end{aligned}
\end{eqnarray}
where $c_1$ is a design constant. Then Eq. (5) becomes 
\begin{eqnarray}
\begin{aligned}
e_{1}(k+1) = c_{1} e_{1}(k) + h e_{2}(k).
\end{aligned}
\end{eqnarray}

\textbf{Step 2.} The final control law is  synthesized to drive $e_2(k)$ towards zero or a small value. The error variable $e_{2}(k)$ as the forward signal is written as
\begin{eqnarray}
\begin{aligned}
M^{+}(k)e_{2}(k+1) &= M^{+}(k)x_{2}(k+1) - M^{+}(k) \alpha(k+1) \\
&= u(k) - g(k) - M^{+}(k) \alpha(k+1) - \tau_{d}(k),
\end{aligned}
\end{eqnarray}
where $\alpha(k+1)$ is written as:
\begin{eqnarray}
\begin{aligned}
\alpha(k+1) &= \frac{1}{h} (c_{1} (c_{1} e_{1}(k) + h e_{2}(k)) \\
&\quad - h x_{2}(k) - x_{1}(k) + x_{1 \mathrm {d}}(k+2)).
\end{aligned}
\end{eqnarray}

The control input $u(k)$ is selected as
\begin{eqnarray}
u(k)=\hat{f}(k) +c_{2} e_{2}(k)
\end{eqnarray}
where $c_2$ is a design constant, $\hat{f}(k)$ is an estimate of the combined unknown dynamics $f(k)$ by neural networks, and $f(k)$ is written as
\begin{eqnarray}
f(k) = g(k) + M^{+}(k) \alpha(k+1).
\end{eqnarray}

Substituting Eq. (11) and Eq. (12) into Eq. (9) yields
\begin{eqnarray}
M^{+}(k)e_{2}(k+1)=c_{2} e_{2}(k) + \tilde{f}(k) - \tau_{d}(k),
\end{eqnarray}
where 
\begin{eqnarray}
\tilde{f}(k) = \hat{f}(k) - f(k).
\end{eqnarray}

\begin{figure}[ht]
	\centering
	\includegraphics[scale=0.36]{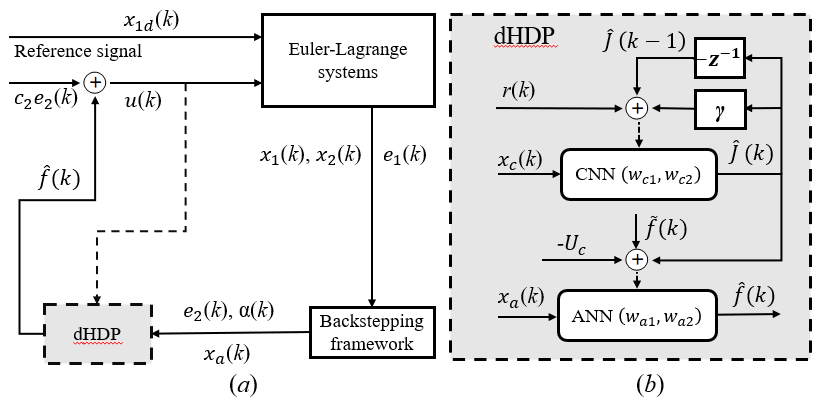}
	\caption{Schematic diagram of the proposed tracking control framework.}
	\label{fig:labe1}
\end{figure}

\section{Reinforcement Learning To Provide Data-driven Feed-forward Input Control}
The dHDP block is used to provide control input for unaddressed dynamics due to a lack of a mathematical description of system Eq. (1). Here we use dHDP as a basic structural framework to approximate the cost-to-go function and the optimal control policy at the same time \cite{si2001online,si2004handbook}. The dHDP has been shown a feasible tool for solving complex and realistic problems including the stabilization, tracking, reconfiguring control of Apache helicopters \cite{enns2002apache,enns2003helicopterj, enns2003helicopter}, damping low frequency oscillations in large power systems \cite{lu2008direct}, and wearable robots with human in the loop \cite{wen2017new, wen2019online, zhang2019adaptive}. We therefore consider the dHDP can potentially provide the necessary online learning capability in the backstepping design and together, we provide a new learning control method that improves the reproducibility of results when applied to meaningful real applications.  

\subsection{Basic Formulation}
The dHDP based reinforcement learning feed-forward control $\hat{f}(k)$ is as shown in Fig. 1. The stage cost $r(k)$ is defined as
\begin{eqnarray}
r(k) = e_{1}^{T}(k) Q e_{1}(k) +  u^{T}(k) R u(k),
\end{eqnarray}
where $Q \in \mathbb{R}^{n\times n}$, $R \in \mathbb{R}^{n\times n}$ are positive semi-deﬁnite matrices. Then, the cost-to-go $J(k)$ is written as
\begin{eqnarray}
J(k) = \sum_{j=1}^{\infty} \gamma^{j} r(k+j),
\end{eqnarray}
where $0 < \gamma < 1$ is a discount factor for the infinite-horizon tracking problem. We require $r(k)$ to be a semi-definite function of the output error $e_1(k)$ and control $u(k)$, so the cost function is well-defined. Based on Eq. (16), we formulate the following Bellman equation:
\begin{eqnarray}
J(k-1) = \gamma J(k) + r(k).
\end{eqnarray}

\subsection{Actor-Critic Networks}
The dHDP design follows that in \cite{si2001online} with an actor neural network and a critic neural network. Hyperbolic tangent is used as the transfer function in the actor-critic networks to approximate the control policy and the cost-to-go function.
\subsubsection{Critic Neural Network}
The critic neural network (CNN) consists of three layers of neurons, namely the input layer, the hidden layer and the output layer. The input and output of CNN are
\begin{eqnarray}
x_c(k) = [x_{a}(k), u(k)]^{T},
\end{eqnarray}
\begin{eqnarray}
\hat{J}(k) = \hat{w}_{c 2}(k) * \phi(\hat{w}_{c 1}(k) * x_c(k)) = \hat{w}_{c2}\phi_{c}(k),
\end{eqnarray}
where 
\begin{eqnarray}
x_a(k) = [x_1(k), x_2(k), e_1(k), e_2(k), x_{\operatorname{1d}}(k+2)]^T,
\end{eqnarray}
In the above, $\hat{w}_{c 1}$ and $\hat{w}_{c 2}$ are the estimated weight matrices between the input and hidden, and output layers, respectively. $\phi(\cdot)$
is the hyperbolic tangent activation function,
\begin{eqnarray}
\phi(v)=\frac{1-\exp (-v)}{1+\exp (-v)}.
\end{eqnarray}

From Eq. (17), the prediction error $e_{c}(k)$ is formulated as
\begin{eqnarray}
e_{c}(k) = \gamma \hat{J}(k) - \left[\hat{J}(k - 1) - r(k)\right].
\end{eqnarray}
The weights of CNN are updated to minimize the following approximation cost
\begin{eqnarray}
E_{c}(k) = {1\over 2}e_{c}^{T}(k) e_{c} (k).
\end{eqnarray}

Gradient descent is used to adjust the critic weights. For the input-to hidden layer,
\begin{eqnarray}
\begin{aligned}
&\Delta \hat{w}_{c1}(k)=l_{c}\left[-\frac{\partial E_{c}(k)}{\partial \hat{w}_{c1}(k)}\right] \\
&\frac{\partial E_{c}(k)}{\partial \hat{w}_{c1}(k)} =\frac{\partial E_{c}(k)}{\partial \hat{J}(k)} \frac{\partial \hat{J}(k)}{\partial \phi_{c}(k)} \frac{\partial \phi_{c}(k)}{\partial v_{c}(k)} \frac{\partial v_{c}(k)}{\partial \hat{w}_{c1}(k)} \\
&\quad \quad \quad \quad =\gamma e_{c}(k) \hat{w}_{c2}(k)\left[\frac{1}{2}\left(1-\phi_{c}^{2}(k)\right)\right] x_c(k),
\end{aligned}
\end{eqnarray}
Similarly for the hidden-to-output layer,
\begin{eqnarray}
\begin{aligned}
&\Delta \hat{w}_{c2}(k)=l_{c}\left[-\frac{\partial E_{c}(k)}{\partial \hat{w}_{c2}(k)}\right] \\
&\frac{\partial E_{c}(k)}{\partial \hat{w}_{c2}(k)}=\frac{\partial E_{c}(k)}{\partial \hat{J}(k)} \frac{\partial \hat{J}(k)}{\partial \hat{w}_{c2}(k)}=\gamma e_{c}(k) \phi_{c}(k).
\end{aligned}
\end{eqnarray}

In the above, $l_{c}>0$ is the learning rate.
\subsubsection{Action Neural Network}
In this algorithm, the action neural network (ANN) is to approximate the unknown dynamics $f(k)$ in Eq. (12). The input to the ANN is $x_a(k)$, the respective output is given as follows:
\begin{eqnarray}
\hat{f}(k) = \hat{w}_{a2}(k) * \phi(\hat{w}_{a 1}(k) * x_{a}(k)) = \hat{w}_{a2}\phi_{a}(k),
\end{eqnarray}
where $\hat{w}_{a 1}$ and $\hat{w}_{a 2}$ are the estimated weight matrices. 

The ANN weights are adjusted to minimize the following cost,
\begin{eqnarray}
E_{a}(k) = {1\over 2}e_{a}^{T}(k)e_{a}(k),
\end{eqnarray}
where
\begin{eqnarray}
e_{a}(k) = \hat{J}(k) + \Vert \hat{f}(k) \Vert - U_c,
\end{eqnarray}
In the above, $U_c$ is the ultimate performance objective in the tracking control design paradigm, which is defined as $U_c = 0$ under the current problem formulation; $\tilde{f}(k) = \hat{f}(k)-f(k)$ is defined in Eq. (14). The desired tracking performance will be achieved if $\|\tilde{f}(k)\|$ approaches 0.

\textbf{Remark 1.} From Eq. (13), we have that
\begin{eqnarray}
\tilde{f}(k) = M^{+}(k)e_2(k + 1) - c_2e_2(k) + \tau_{d}(k).
\end{eqnarray}
To compute $e_{a}(k)$ in Eq. (28), we use an initial estimate $\hat{M}^{+}$ in place of $M^{+}(k)$ in Eq. (29). In the error estimation process, the disturbance $\tau_{d}(k)$ is zero as the feed-forward controller aims at learning the unknown system dynamics.

The weight update rule is again based on gradient descent. For the input-to hidden layer,
\begin{eqnarray}
\begin{aligned}
&\Delta \hat{w}_{a1}(k)=l_{a}\left[-\frac{\partial E_{a}(k)}{\partial \hat{w}_{a1}(k)}\right] \\
&\frac{\partial E_{a}(k)}{\partial \hat{w}_{a1}(k)}= \frac{\partial E_{a}(k)}{\partial \hat{J}(k)}\frac{\partial \hat{J}(k)}{\partial u(k)} \frac{\partial u(k)}{\partial \hat{f}(k)} \frac{\partial f(k)}{\partial \phi_{a}(k)} \frac{\partial \phi_{a}(k)}{\partial v_{a}(k)} \frac{\partial v_{a}(k)}{\partial \hat{w}_{a1}(k)} \\
&\quad \quad \quad ~~ = e_a(k) \left[\hat{w}_{c2}(k) \frac{1}{2}\left(1-\phi_{c}^{2}(k)\right) \hat{w}_{cu}(k)\right] \\
&\quad \quad \quad  \quad~~  \times \hat{w}_{a2}(k) \frac{1}{2}\left(1-\phi_{a}^{2}(k)\right) x_{a}(k),
\end{aligned}
\end{eqnarray}
and for the hidden-to-output layer,
\begin{eqnarray}
\begin{aligned}
	\Delta \hat{w}_{a2}(k) &=l_{a}\left[-\frac{\partial E_{a}(k)}{\partial \hat{w}_{a2}(k)}\right] \\
	\frac{\partial E_{a}(k)}{\partial \hat{w}_{a2}(k)} 
	&=\frac{\partial E_{a}(k)}{\partial \hat{J}(k)} \frac{\partial \hat{J}(k)}{\partial u(k)} \frac{\partial u(k)}{\partial \hat{f}(k)} \frac{\partial \hat{f}(k)}{\partial \hat{w}_{a2}(k)}\\
	&=e_{a}(k) \left[\hat{w}_{c2}(k) \frac{1}{2}\left(1-\phi_{c}^{2}(k)\right) w_{cu}(k)\right] \phi_{a}(k),
\end{aligned}
\end{eqnarray}
where $w_{cu}(k)$ is the weight associated with the input element from ANN, i.e., the part of $w_{c1}$ which connect with $u(k)$, and $l_{a}>0$ is the learning rate.

Algorithm 1 summarizes the implementation procedure of the dHDP-based tracking control.
\begin{center}
	\begin{tabular}{l} %l(left)居左显示 r(right)居右显示 c居中显示
		\hline 
		\textbf{Algorithm 1.} Direct signal tracking control based on dHDP\\
		\hline  
		Specify desired trajectory $x_{1\rm{d}}$; \\
		Initialization: $w_{a}(0)$, $w_{c}(0)$, $x_{1}(0)$, $x_{2}(0)$, $\hat{M}^{+}$;\\
		Set hyperparameters : $l_a$, $l_c$, $c_1$, $c_2$;\\
		Calculate virtual control $\alpha(0)$ according to Eq. (7);\\
		Calculate $e_2(0)$ according to Eq. (6);\\
		Calculate $r(0)$ according to Eq. (15);\\
		\quad\quad\textbf{Backstepping design:}\\
		\quad\quad\quad\quad Calculate $\hat{f}(k)$ according to Eq. (26);\\
		\quad\quad\quad\quad Calculate $u(k)$ according to Eq. (11);\\
		\quad\quad\quad\quad Take control input $u(k)$ into Eq. (2); \\
		\quad\quad\quad\quad Produce $x(k+1)$, $r(k+1)$ according to Eq. (2), (15);\\ 
		\quad\quad\textbf{dHDP design:}\\
		\quad\quad\quad\quad  Obtain $\alpha(k+1)$ by Eq. (7);\\
		\quad\quad\quad\quad  Calculate $e_2(k+1)$ according to Eq. (6);\\
	    \quad\quad\quad\quad  Calculate $\hat{J}(k)$ according to Eq. (19); \\
	    \quad\quad\quad\quad  Calculate $e_{c}(k)$ according to equation Eq. (22); \\
	    \quad\quad\quad\quad\quad\quad  $w_c(k+1) = w_c(k) + \Delta w_c(k)$; \\
	    \quad\quad\quad\quad  Calculate $e_{a}(k)$ (Remark 1) ; \\
    	\quad\quad\quad\quad\quad\quad  $w_a(k+1) = w_a(k) + \Delta w_a(k)$; \\
	    \quad\quad Iterate until converge. \\
		\hline 
	\end{tabular}
\end{center}
\section{Lyapunov Stability Analysis}
In this section, we provide a theoretical analysis for the stability of the overall dynamic system, weight convergence of the actor and critic neural networks, and the Bellman (sub)optimality of the control policy.
\subsection{Preliminaries}
Let $w_{c}^{*}$, $w_{a}^{*}$ denote the optimal weights, that is,
\begin{eqnarray}
\begin{aligned}
w_{a}^{*}=&\arg \min _{\hat{w}_{a}}\Vert\hat{J}(k) + \bar{f}(k) - U_c\Vert, \\
w_{c}^{*}=&\arg \min _{\hat{w}_{c}}\Vert\gamma \hat{J}(k)+r(k)-\hat{J}(k-1)\Vert.
\end{aligned}
\end{eqnarray}
Then, the optimal cost-to-go $J^{*}(k)$ and unknown dynamics $f(k)$ can be expressed as
\begin{eqnarray}
J^{*}(k) = w_{c2}^{*} \phi_{c}(k)+\epsilon_{c}(k), \quad f(k) = w_{a2}^{*} \phi_{a}(k)+\epsilon_{a}(k) 
\end{eqnarray}
where $\epsilon_{c}(k)$ and $\epsilon_{a}(k)$ are the reconstruction errors of the actor and critic neural networks, respectively.

\textbf{Assumption 3.} The optimal weights for the actor-critic networks exist and they are bounded by two positive constants $w_{am}$ and $w_{cm}$, respectively,
\begin{eqnarray}
\left\|w_{a}^{*}\right\| \leq w_{am}, \quad \left\|w_{c}^{*}\right\| \leq w_{cm}.	
\end{eqnarray}
Then, the weight estimation errors of the actor and critic neural networks are described respectively as
\begin{eqnarray}
\tilde{w}_{a}(k):=\hat{w}_{a}(k)-w^{*}_{a},\quad  \tilde{w}_{c}(k):=\hat{w}_{c}(k)-w^{*}_{c}.
\end{eqnarray} 

\textbf{Remark 2.} A weight parameter convergence result was obtained for the dHDP in \cite{liu2012boundedness} under the condition that the weights between the input and hidden layers remain unchanged during learning. The result was later extended to allowing for all the weights in the actor and critic networks to adapt during learning \cite{sokolov2015complete}. Another study \cite{yang2009direct} addressed tracking control using dHDP for a Brunovsky canonical system. Such a system may be mathematically interesting but practically limiting. Additionally, the design in \cite{yang2009direct} requires reference models. In this study, we take reference of \cite{liu2012boundedness} and \cite{sokolov2015complete} to prove our new results on weight convergence for tracking control. Note that \cite{liu2012boundedness} and \cite{sokolov2015complete} are about regulation control, not tracking control. Notice also that both works lack a system stability result.

\textbf{Lemma 1.} Under Assumption 3, consider the weight vector of the hidden-to-output layer in CNN. Let
\begin{eqnarray}
L_1(k)=\frac{1}{l_c} tr \left((\tilde{w}_{c2}(k))^T \tilde{w}_{c2}(k)\right).
\end{eqnarray}
Then its first difference is given by
\begin{eqnarray}
\begin{aligned}
\Delta L_{1}(k)&= -\gamma^{2}\left\|\zeta_{c}(k)\right\|^{2}-\left(1-\gamma^{2} l_{c}\|\phi_{c}(k)\right\|^{2}) \\
& \times\|\gamma \hat{w}_{c2}(k) \phi_{c}(k)+r(k)-\hat{w}_{c2}(k-1) \phi_{c}(k-1)\|^{2} \\
&  +\|\gamma w_{c2}^{*} \phi_{c}(k)+r(k)-\hat{w}_{c2}(k-1) \phi_{c}(k-1)\|^{2},
\end{aligned}
\end{eqnarray}
where $\zeta_{c}(k)=\tilde{w}_{c2}(k) \phi_{c}(k)$ is an approximation error of the critic output.

\textbf{Proof of Lemma 1}. The first difference of $L_1(k)$ can be written as
\begin{eqnarray}
\begin{aligned}
\Delta L_1(k) &= \frac{1}{l_c}tr\left[(\tilde{w}_{c2}(k+1))^{T} \tilde{w}_{c2}(k+1) \right.\\
&\left. - (\tilde{w}_{c2}(k))^{T} \tilde{w}_{c2}(k)\right].
\end{aligned}
\end{eqnarray}
With the updating rule in Eq. (25), $\tilde{w}_{c2}(k+1)$ can be rewritten as
\begin{eqnarray}
\begin{aligned}
\tilde{w}_{c2}(k+1)=& \hat{w}_{c2}(k+1)-w_{c2}^{*} \\
=& \tilde{w}_{c2}(k)-\gamma l_{c} \phi_{c}(k)\left[\gamma \hat{w}_{c2}(k) \phi_{c}(k)\right.\\
&\left.+r(k)-\hat{w}_{c2}(k-1) \phi_{c}(k-1)\right]^{T}.
\end{aligned}
\end{eqnarray}
Then, the first term in the brackets of Eq. (38) can be given as
\begin{eqnarray}
\begin{aligned}
&tr\left[(\tilde{w}_{c2}(k+1))^{T} \tilde{w}_{c2}(k+1)\right] \\
=&(\tilde{w}_{c2}(k))^{T} \tilde{w}_{c2}(k)-2 \gamma l_{c} \tilde{w}_{c2}(k) \phi_{c}(k) \\
&\times\left[\gamma \hat{w}_{c2}(k) \phi_{c}(k)+r(k)-\hat{w}_{c2}(k-1) \phi_{c}(k-1)\right]^{T} \\
&+ \gamma^{2} l_{c}^{2}\left\|\phi_{c}(k)\right\|^{2} \| \gamma \hat{w}_{c2}(k) \phi_{c}(k) + r(k) \\
&- \hat{w}_{c2}(k-1) \phi_{c}(k-1) \|^{2}.
\end{aligned}
\end{eqnarray}
As $\tilde{w}_{c2}(k) \phi_{c}(k)$ is a scalar, by using Eq. (35), we can rewrite the middle term in the above formula as follows:
\begin{eqnarray}
\begin{aligned}
-&2 \gamma l_{c} \tilde{w}_{c2}(k) \phi_{c}(k)\left[\gamma \hat{w}_{c2}(k) \phi_{c}(k)+r(k)\right.\\
&\quad \left.-\hat{w}_{c2}(k-1) \phi_{c}(k-1)\right] \\
=& l_{c}\left(\| \gamma \hat{w}_{c2}(k) \phi_{c}(k)+r(k)-\hat{w}_{c2}(k-1) \phi_{c}(k-1)\right.\\
&-\gamma \tilde{w}_{c2}(k) \phi_{c}(k)\left\|^{2}-\right\| \gamma \tilde{w}_{c2}(k) \phi_{c}(k) \|^{2} \\
&-\left.\|\gamma \hat{w}_{c2}(k) \phi_{c}(k)+r(k)-\hat{w}_{c2}(k-1) \phi_{c}(k-1)\|^{2}\right) \\
=& l_{c}\left(\|\gamma w_{c2}^{*} \phi_{c}(k)+r(k)-\hat{w}_{c2}(k-1) \phi_{c}(k-1)\|^{2}\right.\\
&-\gamma^{2}\left\|\zeta_{c}(k)\right\|^{2} -\| \gamma \hat{w}_{c2}(k) \phi_{c}(k)+r(k) \\
&\left.-\hat{w}_{c2}(k-1) \phi_{c}(k-1) \|^{2}\right).
\end{aligned}
\end{eqnarray}

Substituting Eq. (40) and Eq. (41) into Eq. (38), we obtain Lemma 1. $\hfill\blacksquare$ 

\textbf{Lemma 2.} Under Assumption 3, consider the weight vector of the hidden-to-output layer in ANN. Let
\begin{eqnarray}
L_{2}(k)= \frac{1}{l_{a} \beta_{1}} tr\left[\left(\tilde{w}_{a2}(k)\right)^{T} \tilde{w}_{a2}(k)\right].
\end{eqnarray}
Then its first difference is bounded by
\begin{eqnarray}
\begin{aligned} 
\Delta L_{2}(k) \leq & \frac{1}{\beta_{1}}\left(-(1-l_{a}\|\phi_{a}(k)\|^{2}\|\hat{w}_{c2}(k) C(k)\|^{2})\right.\\ 
&\left. \times\|\hat{w}_{c2}(k) \phi_{c}(k) + d\|^{2}  + 8 \|\zeta_{c}(k)\|^{2} \right.\\
&\left. + 8\|w_{c2}^{*} \phi_{c}(k)\|^{2} + 8 \|\zeta_{a}(k)\|^{2} + 4d^{2} \right. \\
&\left. + \|\hat{w}_{c2}(k) C(k) \zeta_{a}(k)\|^{2}\right),
\end{aligned}
\end{eqnarray}
where $\zeta_{a}(k)=\tilde{w}_{a2}(k) \phi_{a}(k)$ is an approximation error of the action network output; $C(k)=\frac{1}{2}\left(1-\phi_{c}^{2}(k)\right) w_{cu}(k)$; $\beta_{1}>0$ is a weighting factor; 
and the lumped disturbance $d = \|(\hat{M}^{+}-M^{+})e_{2}(k+1) + \zeta_{a}(k) -\tau_{d}(k) - \epsilon_{a}(k)\|$.

\textbf{Proof of Lemma 2.} The first difference of $L_2(k)$ can be written as
\begin{eqnarray}
\begin{aligned}
\Delta L_2(k) &= \frac{1}{l_{a} \beta_{1}}tr\left[(\tilde{w}_{a2}(k+1))^{T} \tilde{w}_{a2}(k+1) \right.\\
&\left. - (\tilde{w}_{a2}(k))^{T} \tilde{w}_{a2}(k)\right].
\end{aligned}
\end{eqnarray}
With the updating rule in Eq. (31), $\tilde{w}_{a2}(k+1)$ can be rewritten as
\begin{eqnarray}
\begin{aligned}
\tilde{w}_{a2}(k+1)=&\hat{w}_{a2}(k+1)-w_{a2}^{*} \\
=& \hat{w}_{a2}(k)-l_{a} \phi_{a}(k) \hat{w}_{c2}(k) C(k)   \\
&  \times[\hat{w}_{c2}(k) \phi_{c}(k) + d ]^{T} - w_{a2}^{*} \\
=& \tilde{w}_{a2}(k)-l_{a} \phi_{a}(k) \hat{w}_{c2}(k) C(k) \\
& \times [\hat{w}_{c2}(k) \phi_{c}(k) + d ]^{T}.
\end{aligned}
\end{eqnarray}
Based on this expression, it is easy to obtain that 
\begin{eqnarray}
\begin{aligned}
&tr \left[\left(\tilde{w}_{a2}(k+1)\right)^{T} \tilde{w}_{a2}(k+1)\right] \\
=&(\tilde{w}_{a2}(k))^{T} \tilde{w}_{a2}(k)+l_{a}^{2}\|\phi_{a}(k)\|^{2}\|\hat{w}_{c2}(k) C(k)\|^{2}\\
&\times\|\hat{w}_{c2}(k) \phi_{c}(k) + d\|^{2} \\
& -2 l_{a} \hat{w}_{c2}(k) C(k) [\hat{w}_{c2}(k) \phi_{c}(k) + d]^{T} \zeta_{a}(k).
\end{aligned}
\end{eqnarray}

Substituting Eq. (46) into Eq. (44), we have
\begin{eqnarray}
\begin{aligned} 
\Delta L_{2}(k)=& \frac{1}{\beta_{1}}(l_{a}\|\phi_{a}(k)\|^{2}\|\hat{w}_{c2}(k) C(k)\|^{2}\\
& \times \|\hat{w}_{c2}(k) \phi_{c}(k) + d\|^{2} \\
& -2 \hat{w}_{c2}(k) C(k)[\hat{w}_{c2}(k) \phi_{c}(k) + d]^{T} \zeta_{a}(k)).
\end{aligned}
\end{eqnarray}
The second term in Eq. (46) can be given as
\begin{eqnarray}
\begin{aligned}
&-2 \hat{w}_{c2}(k) C(k)[\hat{w}_{c2}(k) \phi_{c}(k) + d]^{T} \zeta_{a}(k) \\
=&  \|\hat{w}_{c2}(k) \phi_{c}(k) + d - \hat{w}_{c2}(k) C(k) \zeta_{a}(k)\|^{2}\\ 
& + \|\hat{w}_{c2}(k) C(k) \zeta_{a}(k)\|^{2} - \|\hat{w}_{c2}(k) \phi_{c}(k) + d\|^{2} \\ 
\leq& 2 \|\hat{w}_{c2}(k) \phi_{c}(k) + d\|^{2} + \|\hat{w}_{c2}(k) C(k) \zeta_{a}(k)\|^{2} \\
& - \|\hat{w}_{c2}(k) \phi_{c}(k) + d\|^{2}\\
\leq& 4\|\hat{w}_{c2}(k) \phi_{c}(k)\|^{2} + 4d^{2} + \|\hat{w}_{c2}(k) C(k) \zeta_{a}(k)\|^{2}\\
& - \|\hat{w}_{c2}(k) \phi_{c}(k) + d(k)\|^{2}\\
\leq & 4\|(\tilde{w}_{c2}(k)+w_{c2}^{*}) \phi_{c}(k)\|^{2} + 4 d^{2} + \|\hat{w}_{c2}(k) C(k) \zeta_{a}(k)\|^{2}\\
& - \|\hat{w}_{c2}(k) \phi_{c}(k) + d\|^{2}\\
\leq& 8\|\zeta_{c}(k)\|^{2} + 8\|w_{c2}^{*} \phi_{c}(k)\|^{2} + 4d^{2} +\|\hat{w}_{c2}(k) C(k) \zeta_{a}(k)\|^{2} \\
& - \|\hat{w}_{c2}(k) \phi_{c}(k) + d\|^{2}.
\end{aligned}
\end{eqnarray}
We have thus obtained Lemma 2. $\hfill\blacksquare$

\textbf{Lemma 3.}  Under Assumption 3, consider the weight vector of the input-to-hidden layer in CNN. Let
\begin{eqnarray}
L_{3}(k)=\frac{1}{l_{c} \beta_{2}} tr\left[(\tilde{w}_{c1}(k))^{T} \tilde{w}_{c1}(k)\right].
\end{eqnarray}
Then its first difference is bounded by
\begin{eqnarray}
\begin{aligned}
\Delta L_{3}(k) \leq& \frac{1}{\beta_{2}}\left(\gamma^{2} l_{c} \| \gamma \hat{w}_{c2}(k) \phi_{c}(k)+r(k)\right. \\
&-\hat{w}_{c2}(k-1) \phi_{c}(k-1)\|^{2}\| A(k)\|^{2}\| x_c(k) \|^{2} \\
&+\gamma \|\tilde{w}_{c1}(k) x_c(k) A^{T}(k)\|^{2}+\gamma \| \gamma \hat{w}_{c2}(k) \phi_{c}(k) \\
&\left. + r(k)-\hat{w}_{c2}(k-1) \phi_{c}(k-1) \|^{2}\right),
\end{aligned}
\end{eqnarray}
where $\beta_{2}>0$ is a weighting factor and $A(k)$ is a vector, with $A(k)=$ $\frac{1}{2}\left(1-\phi_{c}^{2}(k)\right) \hat{w}_{c2}(k)$.

\textbf{Proof of Lemma 3.} The first difference of $L_3(k)$ can be written as
\begin{eqnarray}
\begin{aligned}
\Delta L_3(k) &= \frac{1}{l_{c} \beta_{2}}tr\left[(\tilde{w}_{c1}(k+1))^{T} \tilde{w}_{c1}(k+1) \right.\\
&\left. - (\tilde{w}_{c1}(k))^{T} \tilde{w}_{c1}(k)\right].
\end{aligned}
\end{eqnarray}
With the updating rules in Eq. (24), $\hat{w}_{c1}(k+1)$ can be written as 
\begin{eqnarray}
\begin{aligned}
\hat{w}_{c1}(k+1)=&\hat{w}_{c1}(k)-\gamma l_{c}\left(\gamma \hat{w}_{c2}(k) \phi_{c}(k)\right.\\
&\left.+r(k)-\hat{w}_{c2}(k-1) \phi_{c}(k-1)\right)^{T} B(k),
\end{aligned}
\end{eqnarray} 
where $B(k)=\frac{1}{2}\left(1-\phi_{c}^{2}(k)\right) \hat{w}_{c2}(k) x_{c}(k)$. 
Following the same approach as earlier, we can express $\tilde{w}_{c1}(k+1)$ by 
\begin{eqnarray}
\begin{aligned}  
\tilde{w}_{c1}(k+1)=& \hat{w}_{c1}(k+1)-w_{c1}^{*} \\
=& \tilde{w}_{c1}(k)-\gamma l_{c}\left(\gamma \hat{w}_{c2}(k) \phi_{c}(k)+r(k)\right.\\ 
&\left.-\hat{w}_{c2}(k-1) \phi_{c}(k-1)\right)^{T} B(k) .
\end{aligned}
\end{eqnarray}

To facilitate the development, the following notation is introduced: 
\begin{eqnarray}
B^{T}(k) B(k)=x_{c}^{T}(k) A^{T}(k) A(k) x_{c}(k)=\|A(k)\|^{2}\|x_{c}(k)\|^{2}.
\end{eqnarray}
Then, we obtain
\begin{eqnarray}
\begin{aligned}
&tr\left[(\tilde{w}_{c1}(k+1))^{T} \tilde{w}_{c1}(k+1)\right] \\
=&(\tilde{w}_{c1}(k))^{T} \tilde{w}_{c1}(k)+\gamma^{2} l_{c}^{2} \| \gamma \hat{w}_{c2}(k) \phi_{c}(k) \\
&+r(k)-\hat{w}_{c2}(k-1) \phi_{c}(k-1) \|^{2} B^{T}(k) B(k) \\
&-2 \gamma l_{c}\left(\gamma \hat{w}_{c2}(k) \phi_{c}(k) + r(k)\right.\\
&\left.-\hat{w}_{c2}(k-1) \phi_{c}(k-1)\right) B^{T}(k) \tilde{w}_{c1}(k).
\end{aligned}
\end{eqnarray}
By introducing the property of trace function, 
\begin{eqnarray}
tr \left(x_c(k) A^{T}(k) \tilde{w}_{c1}(k)\right)=tr\left(\tilde{w}_{c1}(k) x_c(k) A^{T}(k)\right),
\end{eqnarray}
the last term in (55) can be expressed as
\begin{eqnarray}
\begin{aligned}
&-2 \gamma l_{c}\left(\gamma \hat{w}_{c2}(k) \phi_{c}(k)+r(k) - \hat{w}_{c2}(k-1) \phi_{c}(k-1)\right) \\
& \times x_{c}(k) A^{T}(k) \tilde{w}_{c1}(k)\\
=&\gamma l_{c}\left(\| \gamma \hat{w}_{c2}(k) \phi_{c}(k)+r(k)-\hat{w}_{c2}(k-1) \phi_{c}(k-1)\right. \\
&-\tilde{w}_{c1}(k) x_{c}(k) A^{T}(k)\left\|^{2}-\right\| \tilde{w}_{c1}(k) x_{c}(k) A^{T}(k) \|^{2} \\
&\left.-\|\gamma \hat{w}_{c2}(k) \phi_{c}(k)+r(k)-\hat{w}_{c2}(k-1) \phi_{c}(k-1)\|^{2}\right).
\end{aligned}
\end{eqnarray}

Therefore, substituting Eq. (56) and Eq. (57) into Eq. (52), we have Lemma 3. $\hfill\blacksquare$ 

\textbf{Lemma 4.} Under Assumption 3, consider the weight vector of the input-to-hidden layer in ANN. Let
\begin{eqnarray}
L_{4}(k)= \frac{1}{l_{a} \beta_{3}} tr\left[(\tilde{w}_{a1}(k))^{T} \tilde{w}_{a1}(k)\right],
\end{eqnarray}
Then its first difference is bounded by
\begin{eqnarray}
\begin{aligned} 	
\Delta L_{4}(k) \leq& \frac{1}{\beta_{3}}(l_{a}\|\hat{w}_{c2}(k) \phi_{c}(k) + d(k) \|^{2}  \|x_{a}(k)\|^{2} \\
& \times \|\hat{w}_{c2}(k) C(k) D^{T}(k)\|^{2} + \| \hat{w}_{c2}(k) \phi_{c}(k) + d \|^{2} \\
& + \|\tilde{w}_{a1}(k) x_{a}(k)\|^{2} \|\hat{w}_{c2}(k) C(k) D^{T}(k)\|^{2}),
\end{aligned}
\end{eqnarray}
where $D(k)=\frac{1}{2}\left(1-\phi_{a}^{2}(k)\right) \hat{w}_{a2}(k)$, and $\beta_{3}>0$ is a weighting factor.

\textbf{Proof of Lemma 4.} The first difference of $L_4(k)$ can be written as
\begin{eqnarray}
\begin{aligned}
\Delta L_4(k) &= \frac{1}{l_{a} \beta_{3}}tr\left[(\tilde{w}_{a1}(k+1))^{T} \tilde{w}_{a1}(k+1) \right.\\
&\quad \left. - (\tilde{w}_{a1}(k))^{T} \tilde{w}_{a1}(k)\right].
\end{aligned}
\end{eqnarray}
With the updating rule in Eq. (30), $\tilde{w}_{a1}(k+1)$ can be rewritten as
\begin{eqnarray}
\begin{aligned}
\tilde{w}_{a1}(k+1)=&\hat{w}_{a1}(k+1)-w_{a1}^{*}\\
=&\tilde{w}_{a1}(k) - l_{a} [\hat{w}_{c2}(k) \phi_{c}(k) + d] \\
&\times D(k) C^{T}(k)(\hat{w}_{c2}(k))^{T} x_{a}^{T}(k).
\end{aligned}
\end{eqnarray}

Let us consider
\begin{eqnarray}
\begin{aligned}
&tr\left[(\tilde{w}_{a1}(k+1))^{T} \tilde{w}_{a1}(k+1)\right] \\
&=(\tilde{w}_{a1}(k))^{T} \tilde{w}_{a1}(k) + l_{a}^{2}\|\hat{w}_{c2}(k) \phi_{c}(k) + d \|^{2}\\
&\quad  \times \|\hat{w}_{c2}(k) C(k) D^{T}(k)\|^{2}\|x_{a}(k)\|^{2} \\
&\quad - 2 l_{a} \hat{w}_{c2}(k) C(k) D^{T}(k) (\hat{w}_{c2}(k) \phi_{c}(k) + d)^{T}\tilde{w}_{a1}(k) x_{a}(k). \\
\end{aligned}
\end{eqnarray}
Then, by using the property of trace function $tr(X^{T}Y + Y^{T}X) = tr(X^{T}Y) + tr([X^{T}Y]^{T}) = 2tr(X^{T}Y)$ and $tr(XY) = tr(YX)$, the last term in Eq. (62) is bounded by 
\begin{eqnarray}
\begin{aligned}
&-2 l_{a} \hat{w}_{c2}(k) C(k) D^{T}(k) (\hat{w}_{c2}(k) \phi_{c}(k) + d)^{T} \tilde{w}_{a1}(k) x_{a}(k) \\
\leq& l_{a}\left(\| \hat{w}_{c2}(k) \phi_{c}(k) + d \|^{2}+ \|\tilde{w}_{a1}(k) x_{a}(k)\|^{2} \right.\\
&\left. \times \|\hat{w}_{c2}(k) C(k) D^{T}(k)\|^{2} \right).
\end{aligned}
\end{eqnarray}

We have the statement of Lemma 4 by substituting Eq. (61) and Eq. (62) into Eq. (60).  $\hfill\blacksquare$

\subsection{Stability, convergence, and (sub)optimality results}
With Lemmas 1-4 in place, we are now in a position to provide results on closed-loop stability of the system, convergences of the neural networks in dHDP, and the Bellman (sub)optimality of the resulted control policy. 

\textbf{Definition 1.}~(Uniformly Ultimately Boundedness of a discrete time dynamical system \cite{michel2008stability, sarangapani2018neural})~A dynamical system is said to be uniformly ultimately bounded with ultimate bound $b>0,$ if for any $a>0$ and $k_{0}>0,$ there exists a positive number $N=N(a, b)$ independent of $k_{0}$, such that $\|\tilde{\xi}(k)\| \leq b$ for all $k \geq N+k_{0}$ whenever $\left\|\tilde{\xi}\left(k_{0}\right)\right\| \leq a$. 

In the following, $\tilde{\xi}(k)$ can represent tracking errors $e_{1}(k)$, $e_{2}(k)$ or weight approximation errors $\tilde{w}_{a}(k)$, $\tilde{w}_{c}(k)$, which are related to the stable system or the weight convergence of the actor and critic neural networks, respectively.

\textbf{Theorem 1.}~(Stability Result)~Let Assumption 2 and Assumption 3 hold, and the tracking errors $e_{1}(k)$ and $e_{2}(k)$ defined in Eq. (4) and Eq. (6), respectively. Then the considered system is uniformly ultimately bounded by their respective initial errors if $c_1$ in Eq. (8) and $c_{2}$ in Eq. (11) satisfy the following condition,
\begin{eqnarray}
\| c_{1} \| < \frac{\sqrt{2}}{2}, \quad  \| c_{2} \| < \frac{1}{2}\sqrt{2 M_{min} - 4h^2} 
\end{eqnarray}
where $M_{min}$ is given in Eq. (3).

\textbf{Proof of Theorem 1.} We introduce a candidate Lyapunov function:
\begin{eqnarray}
L_{s}(k) = e_{1}^T(k) e_{1}(k) + (M^{-}(k)e_{2}(k))^T(M^{-}(k)e_{2}(k)).
\end{eqnarray}
The first difference of $L_{s}(k)$ is given as
\begin{eqnarray}
\begin{aligned}
\Delta L_{s}(k) =& e_{1}^T(k+1) e_{1}(k+1) \\
&+ (M^{-}(k+1)e_{2}(k+1))^T(M^{-}(k+1)e_{2}(k+1)) \\
& - e_{1}^T(k) e_{1}(k) - (M^{-}(k)e_{2}(k))^T(M^{-}(k)e_{2}(k)),
\end{aligned}
\end{eqnarray}
by substituting $e_{1}(k+1)$ and $M^{-}(k+1)e_{2}(k+1)$ from Eq. (8) and (13), we have
\begin{eqnarray}
\begin{aligned}
\Delta L_{s}(k) =&  (2\| c_1 \|^{2} - 1) e_{1}^T(k) e_{1}(k) \\
&  + (2 \| c_{2}\|^{2} - \|(M^{-}(k))^{T}(M^{-}(k))\| + 2h^{2}) e_{2}^T(k) e_{2}(k) \\
&+ 8 \Vert \zeta_{a}(k) \Vert^{2} + 8\Vert \epsilon_{a}(k) \Vert^{2} + 4\Vert \tau_{d}(k) \Vert^{2}.
\end{aligned}
\end{eqnarray}

Based on Eq. (3), and also let $H_{1} = 8 \Vert \zeta_{a}(k) \Vert^{2} + 8\Vert \epsilon_{a}(k) \Vert^{2} + 4\Vert \tau_{d}(k) \Vert^{2}$, we obtain
\begin{eqnarray}
\begin{aligned} 
\Delta L_{s}(k) <& - (1 - 2\Vert c_1 \Vert^{2}) e_{1}^2(k) \\
&- [M_{min} - 2(\Vert c_{2}\Vert^{2} + h^2)] e_{2}^2(k) + H_{1}.
\end{aligned}
\end{eqnarray}
and $H_{1}$ can be bounded by
\begin{eqnarray}
H_{1} \leq 8 \Vert \zeta_{am} \Vert^{2} + 8\Vert \epsilon_{am} \Vert^{2} + 4\Vert \tau_{dm} \Vert^{2}=H_{1m},
\end{eqnarray}
where $\zeta_{am}$, $\epsilon_{am}$ and $\tau_{dm}$ are the upper bound of $\zeta_{a}(k)$ in Eq. (43), $\epsilon_{a}(k)$ in Eq. (33), and $\tau_{d}(k)$ in Eq. (2), respectively.

Therefore, for $c_1$, $c_2$ chosen from Eq. (64), and 
\begin{eqnarray}
\begin{aligned}
&\Vert e_{1}(k) \Vert >\sqrt{\frac{H_{1m}}{1 - 2\Vert c_1 \Vert^{2}}} \\
~\operatorname{or} ~&\Vert e_{2}(k) \Vert >\sqrt{\frac{H_{1m}}{M_{min} - 2(\Vert c_{2}\Vert^{2} + h^2)}},
\end{aligned}
\end{eqnarray}
the first difference $\Delta L_{s}<0$.

According to Definition 1 and the bounded initial states and weights, this demonstrates that the errors $e_{1}(k)$ and $e_{2}(k)$ are uniformly ultimately bounded from time step $k$ to $k+1$, and the boundness of control input can also be ensured according to Eq. (11).

\textbf{Theorem 2.}~(Weight Convergence) Under Assumption 3 and the stage cost as given in Eq. (15) and based on Theorem 1, let the weights of the actor and critic neural networks be updated according to Eq. (24), (25), (30) and (31), respectively. Then $\tilde{w}_{c}$ and $\tilde{w}_{a}$ are uniformly ultimately bounded provided that the following conditions are met:
\begin{eqnarray}
\begin{aligned} 
&l_{c}<\min _{k} \frac{\beta_{2}-\gamma}{\gamma^{2} \beta_{2}\left(\left\|\phi_{c}(k)\right\|^{2}+\frac{1}{\beta_{2}}\|A(k)\|^{2}\|x_{c}(k)\|^{2}\right)}  \\ 
&l_{a}<\min _{k}\left(\beta_{3}-\beta_{1}\right)\left(\beta_{3}\|\left(\hat{w}_{c2}(k)\right)^{T} C(k)\|^{2}\|\phi_{a}(k)\|^{2}\right.\\
&\quad \quad \left.+\beta_{1}\|\hat{w}_{c2}(k) C(k) D^{T}(k)\|^{2}\|x_{a}(k)\|^{2}\right)^{-1}.
\end{aligned}
\end{eqnarray}

\textbf{Remark 3.} With $\gamma$, $\beta_{1}$, $\beta_{2}$, $\beta_{3}$ provided in Eq. (16), (42), (49) and (58), respectively, we can choose $l_{c}$, $l_{a}$ to satisfy (71) by setting $\beta_{2}>\gamma>0$ and $\beta_{3}>\beta_{1}>0$.

\textbf{Proof of Theorem 2.} We introduce a candidate of Lyapunov function:
\begin{eqnarray}
L_{w}(k) = L_1(k) + L_2(k) + L_3(k) + L_4(k),
\end{eqnarray}
where $L_1(k)$, $L_2(k)$, $L_3(k)$ and $L_4(k)$ are shown in Eq. (36), (42), (49) and (58). Then the first difference of $L_{w}(k)$ is given as
\begin{eqnarray}
\begin{aligned}
\Delta L_{w}(k)\leq& -(\gamma^{2}-\frac{8}{\beta_{1}})\left\|\zeta_{c}(k)\right\|^{2}-(1-\gamma^{2} l_{c}\|\phi_{c}(k)\|^{2}\\
&-\frac{\gamma^{2} l_{c}}{\beta_{2}}\|A(k)\|^{2}\|x_c(k)\|^{2}-\frac{\gamma}{\beta_{2}}) \| \gamma \hat{w}_{c2}(k) \phi_{c}(k) \\
&+r(k) -\hat{w}_{c2}(k-1) \phi_{c}(k-1)\|^{2}- \left\| \hat{w}_{c2}(k) \phi_{c}(k) \right. \\
&\left. +  d\right\|^{2} (\frac{1}{\beta_{1}}-\frac{l_{a}}{\beta_{1}}\|\hat{w}_{c2}(k) C(k)\|^{2}\|\phi_{a}(k)\|^{2} \\
& -\frac{l_{a}}{\beta_{3}}\|\hat{w}_{c2}(k) C(k) D^{T}(k)\|^{2}\|x_{a}(k)\|^{2} - \frac{1}{\beta_{3}})  \\
& + H_{2},
\end{aligned}
\end{eqnarray}
where $H_{2}$ is defined as 
\begin{eqnarray}
\begin{aligned}
H_{2} =& \|\gamma w_{c2}^{*} \phi_{c}(k)+r(k)-\hat{w}_{c2}(k-1) \phi_{c}(k-1)\|^{2} \\
+& \frac{1}{\beta_{1}}\left\{8\|w_{c2}^{*} \phi_{c}(k)\|^{2}+ 4d^{2} +\|\hat{w}_{c2}(k) C(k) \zeta_{a}(k)\|^{2}  \right\} \\
+&   \frac{\gamma}{\beta_{2}}\|\tilde{w}_{c1}(k) x_c(k) A^{T}(k)\|^{2}\\
+&\frac{1}{\beta_{3}}\|\hat{w}_{c2}(k) C(k) D^{T}(k)\|^{2} \|\tilde{w}_{a1}(k) x_{a}(k)\|^{2}.
\end{aligned}
\end{eqnarray}
for $l_{c}$ and $l_{a}$ satisfying Eq. (71) and also by selecting $\beta_{2}>\gamma$ and $\beta_{3}>\beta_{1}$, we obtain
\begin{eqnarray}
\begin{aligned}
\Delta L_{w}(k) \leq &-(\gamma^{2}-\frac{8}{\beta_{1}})\left\|\zeta_{c}(k)\right\|^{2} + H_{2}.
\end{aligned}
\end{eqnarray}

Applying the Cauchy–Schwarz inequality, we have
\begin{eqnarray}
\begin{aligned}
H_{2}\leq&(\frac{8}{\beta_{1}}+4 \gamma^{2}+2)\left(w_{2cm} \phi_{cm}\right)^{2}+4 r_{m}^{2} \\
&+\frac{1}{\beta_{1}}\left\{8 \zeta_{am}^{2}+  8d_{m}^{2} + \left(w_{2cm}C_{m}\zeta_{am}\right)^{2}\right\}  \\
&+\frac{\gamma}{\beta_{2}}\left(w_{1cm}x_{cm}A_{m}\right)^{2}+\frac{1}{\beta_{3}}\left( w_{2cm}C_{m}D_{m}w_{1cm}x_{am}\right)^{2} \\
=&H_{2m},
\end{aligned}
\end{eqnarray}
where $\phi_{cm}$, $r_{m}$, $w_{1cm}$, $w_{2cm}$, $A_{m}$, $C_{m}$, $D_{m}$, $x_{am}$, and  $x_{cm}$ are the upper bounds of $\phi_{c}(k)$, $r(k)$, $w_{c1}(k)$, $w_{c2}(k)$, $A(k)$, $C(k)$, $D(k)$, $x_{a}(k)$, and $x_c(k)$, respectively.

Therefore, if $\gamma^{2}-\frac{8}{\beta_{1}}>0,$ that is, $\beta_{1}>\frac{8}{\gamma^{2}}$ and $\gamma \in(0,1),$ then for $l_{a}$, $l_{c}$ with constraints from (70), and
\begin{eqnarray}
\Vert \zeta_{c}(k) \Vert> \sqrt{\frac{H_{2m}}{\gamma^{2}-\frac{8}{\beta_{1}}}},
\end{eqnarray}
the first difference $\Delta L_{w}(k)<0$ holds. 

From Definition 1, this result means that the estimation errors $\tilde{w}_{c}$ and $\tilde{w}_{a}$ are uniformly ultimately bounded from the time step $k$ to $k+1$, respectively. $\hfill\blacksquare$

\textbf{Remark 4.} Given Assumption 3, and that the initial states and weights are bounded, then the initial stage cost and initial output of actor network are bounded. As the feed-forward input $f(k)$ in Eq. (33) is realized via actor network with bounded optimal weights, we have that the initial approximation error of the actor network is bounded. From Definition 1, $\Delta L_{s}<0$ in Eq. (68) and $\Delta L_{w}<0$ in Eq. (74), the tracking errors $e_{1}(k)$, $e_{2}(k)$ and the estimation errors $\tilde{w}_{a}(k)$ and $\tilde{w}_{c}(k)$ are bounded from step $k$ to the next step $k+1$, and the control law $u(k)$ is bounded from step $k$ to the next step $k+1$ as well. Then the resulted stage cost $r(k+1)$ is bounded. By mathematical induction, we have the tracking error $e_{1}(k)$, $e_{2}(k)$ and the estimation errors $\tilde{w}_{a}(k)$, $\tilde{w}_{c}(k)$ uniformly ultimately bounded.

\textbf{Remark 5.} Results of Theorem 1 and Theorem 2 hold under less restrictive conditions than those in \cite{liu2012boundedness, sokolov2015complete} that require bounded stage cost. We require an initially bounded system state and actor-critic network weights only.

\textbf{Theorem 3.}~((Sub)optimality Result) Under the conditions of Theorem 2, the Bellman optimality is achieved within finite approximation error. Meanwhile, the error between the obtained control law $u(k)$
and optimal control $u^{*}(k)$ is uniformly ultimately bounded.

\textbf{Proof of Theorem 3.} From the approximate cost-to-go in Eq. (20) and the cost-to-go expressed in Eq. (33), we have
\begin{eqnarray}
\begin{aligned}
&\|\hat{J}(k))-J^{*}(k)\| \\
=&\|\hat{w}_{c2}(k) \phi_{c}(k)-w_{c2}^{*} \phi_{c}(k)-\epsilon_{c}(k)\| \leqslant\|\tilde{w}_{c2}(k)\| \phi_{cm}+\epsilon_{cm}.
\end{aligned}
\end{eqnarray}

Similarly, from (11), (26) and (33), we have
\begin{eqnarray}
\|u(k)-u^{*}(k)\| \leqslant\|\tilde{w}_{a2}(k)\| \phi_{a m}+\epsilon_{am},
\end{eqnarray}
where $\phi_{cm}$, $\epsilon_{cm}$ and $\epsilon_{am}$ are the upper bound of $\phi_{c}(k)$, $\epsilon_{c}(k)$ and $\epsilon_{a}(k)$. This comes directly as $\|\tilde{w}_{c2}\|$ and $\|\tilde{w}_{a2}\|$ are both uniformly ultimately bounded as the time step $k$ increases as shown in Theorem 2. It demonstrates that the Bellman optimality is achieved within finite approximation errors. $\hfill\blacksquare$ 

\section{Simulation Study}

We use two examples to demonstrate how the proposed algorithm works and how it improves reproducibility of results over the original dHDP for data-driven tracking control. 

Example 1. We consider a single-link robot manipulator with the following motion equation:
\begin{eqnarray}
M \ddot{q}(t) + G(q(t)) + \tau_{d}(t) =  \tau(t),
\end{eqnarray}
where $G(q(t)) = \frac{1}{2} \times 9.8 \times m \times l\times sin(q(t))$; $m$ and $l$ are the mass and the half length of the manipulator, respectively. The values of $M$, $m$, $l$, $\tau_{d}$ and initial state are different in different simulation cases below (refer to Table \uppercase\expandafter{\romannumeral1}). Note that, the model in Eq. (80) is to provide a simulated environment in place of a real physical environment. That is to say that the proposed approach is data-driven.

The feedback gain parameters $c_{1}$ and $c_{2}$ in Eq. (64) are chosen as $c_{1}= 0.7$ and $c_{2}= -5$. The CNN and ANN in Fig.1 each has six hidden nodes. The discount factor $\gamma$ in Eq. (16) is chosen as 0.95. The continuous time system dynamics in Eq. (80) is discretized by the Runge-Kutta discretization method with $h=0.02s$.

The per sample mean square error (MSE) defined below is used in Algorithm 1 to terminate ANN and CNN weight update procedure, and also, it is used for scheduling learning rates. 
\begin{eqnarray}
\mathrm{MSE}=\frac{1}{n^{+}-n^{-}} \Sigma_{k=n^{-}}^{n^{+}}\left\|e_{1}(k)\right\|^{2},
\end{eqnarray} 
where $(n^{+}-n^{-})$ is the number of data samples between time stamps $n^{-}$ and $n^{+}$.

% After the 500$th$ sample, the MSE of the previous 100 samples is calculated, denoted as MSE$_{100^{-}}$. Let $\eta_1 = 0.1\times10^{-3}$ and $\eta_2 = 0.3\times10^{-3}$. The learning rate $l_{a}$, $l_{c}$ will reduce to half if MSE$_{100^{-}} < \eta_1$, or they will be reset as the initial value if MSE$_{100^{-}} > \eta_2$, or they will not change if $\eta_2<$MSE$_{100^{-}} < \eta_2$.

In the following, we demonstrate the effectiveness of the proposed algorithm by first comparing it with dHDP tracking control without backstepping and then, feedback stabilizing control \cite{krstic1995nonlinear}, where the feedback stabilizing control law is $u(k)=c_{2}e_{2}(k)$.  In all the simulation studies below, a trial consists of 6000 consecutive samples.

\subsection{Tracking by dHDP with and without backstepping}

In this comparison study, we set $x_1(0) = -0.1$, $x_2(0) = 0.1$, $m=1$, $l=1$,  $\hat{M}^{+}=5/h$, and $\tau_d = 0$. A total of 50 trials were conducted to obtain results reported here.  We define a trial a success if the MSE of the last 3000 samples (denoted as MSE$_{3000^{+}}$) is less than the MSE of first 3000 samples (denoted as MSE$_{3000^{-}}$). The dHDP alone reached 14\% success rate under the comparison settings. The MSE$_{3000^{+}}$ of the successful cases is 1.454$\times10^{-1}$. In comparison, the success rate of the proposed algorithm is 100\% and the MSE$_{3000^{+}}$ is 2.135$\times10^{-5}$. Typical tracking trials are shown in Fig. 2  and the weights of actor-critic network are shown in Fig. 3. Performance improvement is apparent.

\begin{table}[ht]
	\centering
	\caption{Comprehensive performance evaluation}
	\begin{tabular}{c|c|c|c|c|c|c}
		\toprule
		case &initial state  &$M$ &$m,l$    &$d$          &success rate   & \# reset \\
		\hline 
		1	&(-0.1, 0.1)    &C &(1,1)  &N/A            &  96\%        &2   \\
		\hline 
		2	&(0.1, -0.1)    &C &(1,1)  &N/A             &  82\%        &16  \\
		\hline 
		3	&(0.2, -0.2)    &C &(1,1)  &N/A             &  80\%        &24  \\
		\hline 
		4	&(-0.2, 0.2)    &C &(1,1)  &N/A             &  94\%        &5   \\
		\hline 
		5	&(-0.1, 0.1)    &C &(1,1)  &Pulse          &  76\%        &28  \\
		\hline 
		6	&(-0.1, 0.1)    &C &(1,1)  &Gaus        &  84\%        &12  \\
		\hline 
		7	&(-0.1, 0.1)    &C &(1,2)  &N/A            &  60\%        &26  \\
		\hline 
		8	&(-0.1, 0.1)    &C &(2,2)  &N/A            &  50\%        &35  \\
		\hline 
		9	&(-0.1, 0.1)   &V &(1,1)  &N/A          &  90\%       &8  \\
		\bottomrule 
	\end{tabular}
\end{table}

\begin{figure}[!htb]
\begin{tabular}{cc}
\begin{minipage}[t]{0.48\linewidth}
    \includegraphics[width = 1\linewidth]{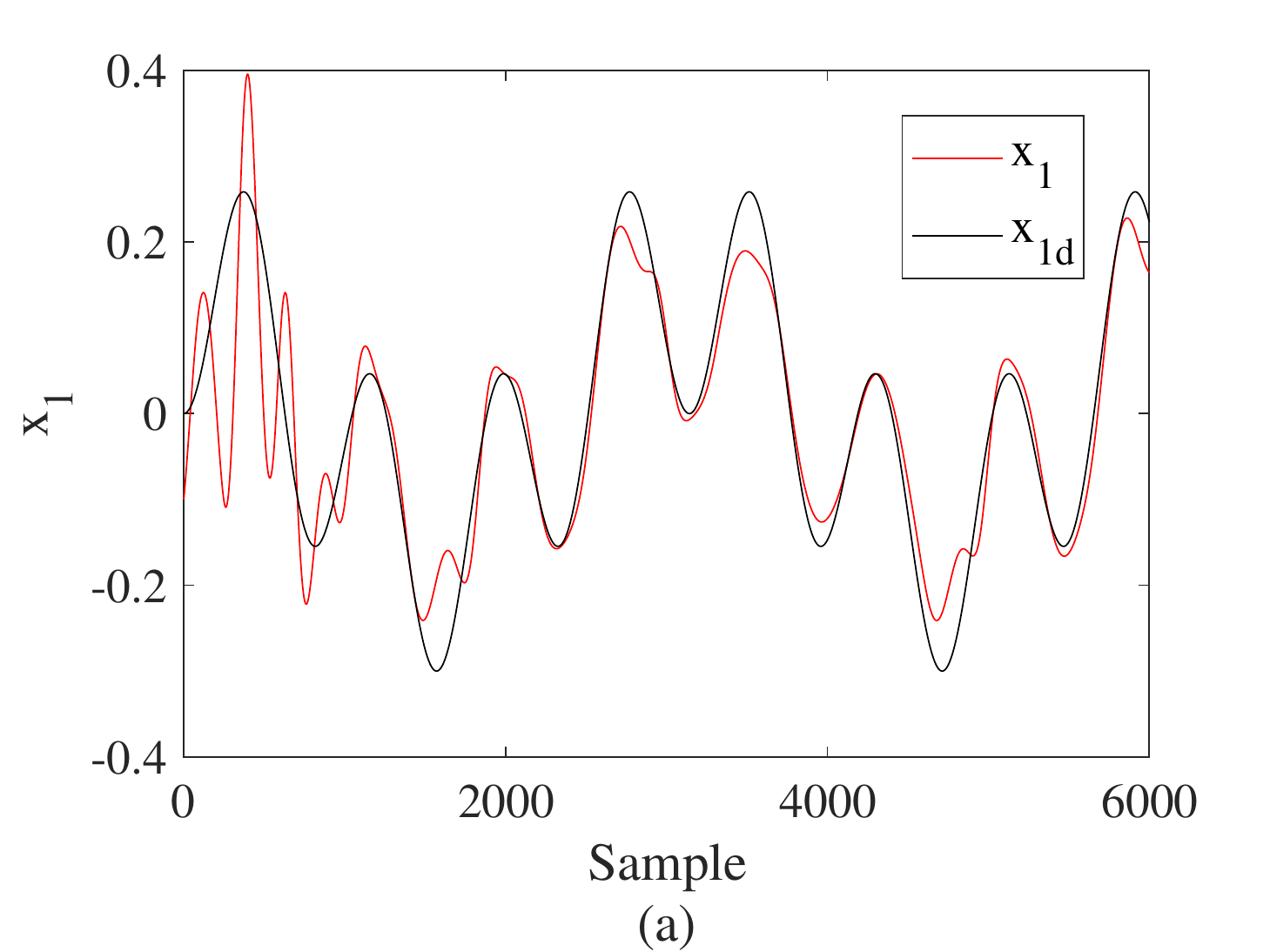}
\end{minipage}
\begin{minipage}[t]{0.48\linewidth}
    \includegraphics[width = 1\linewidth]{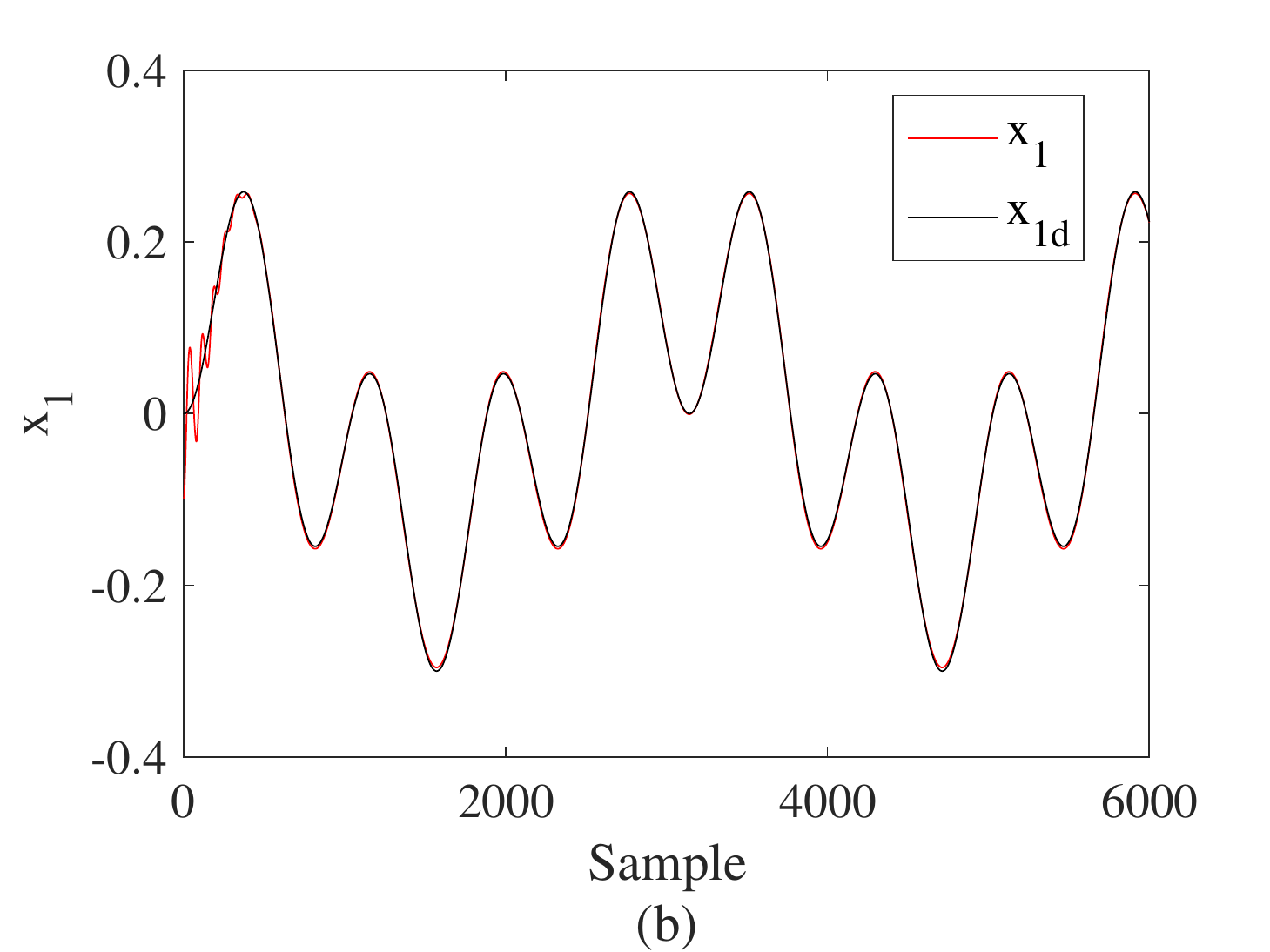}
\end{minipage}
\end{tabular}
\caption{Tracking performance for dHDP without backstepping (a) and tracking performance for dHDP with backstepping (b).}
\end{figure}

\begin{figure}[!htb]
\begin{tabular}{cc}
\begin{minipage}[t]{0.48\linewidth}
    \includegraphics[width = 1\linewidth]{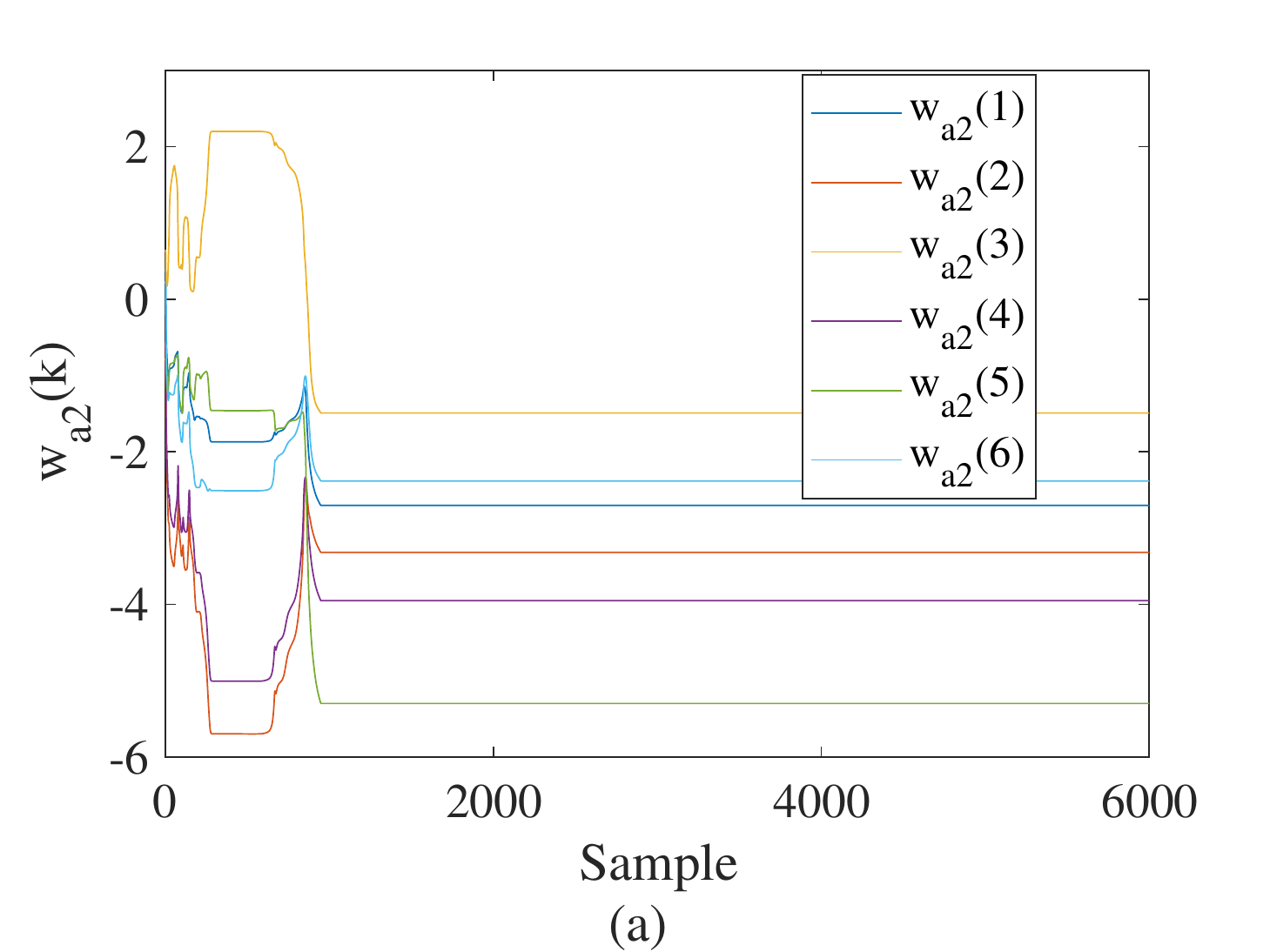}
\end{minipage}
\begin{minipage}[t]{0.48\linewidth}
    \includegraphics[width = 1\linewidth]{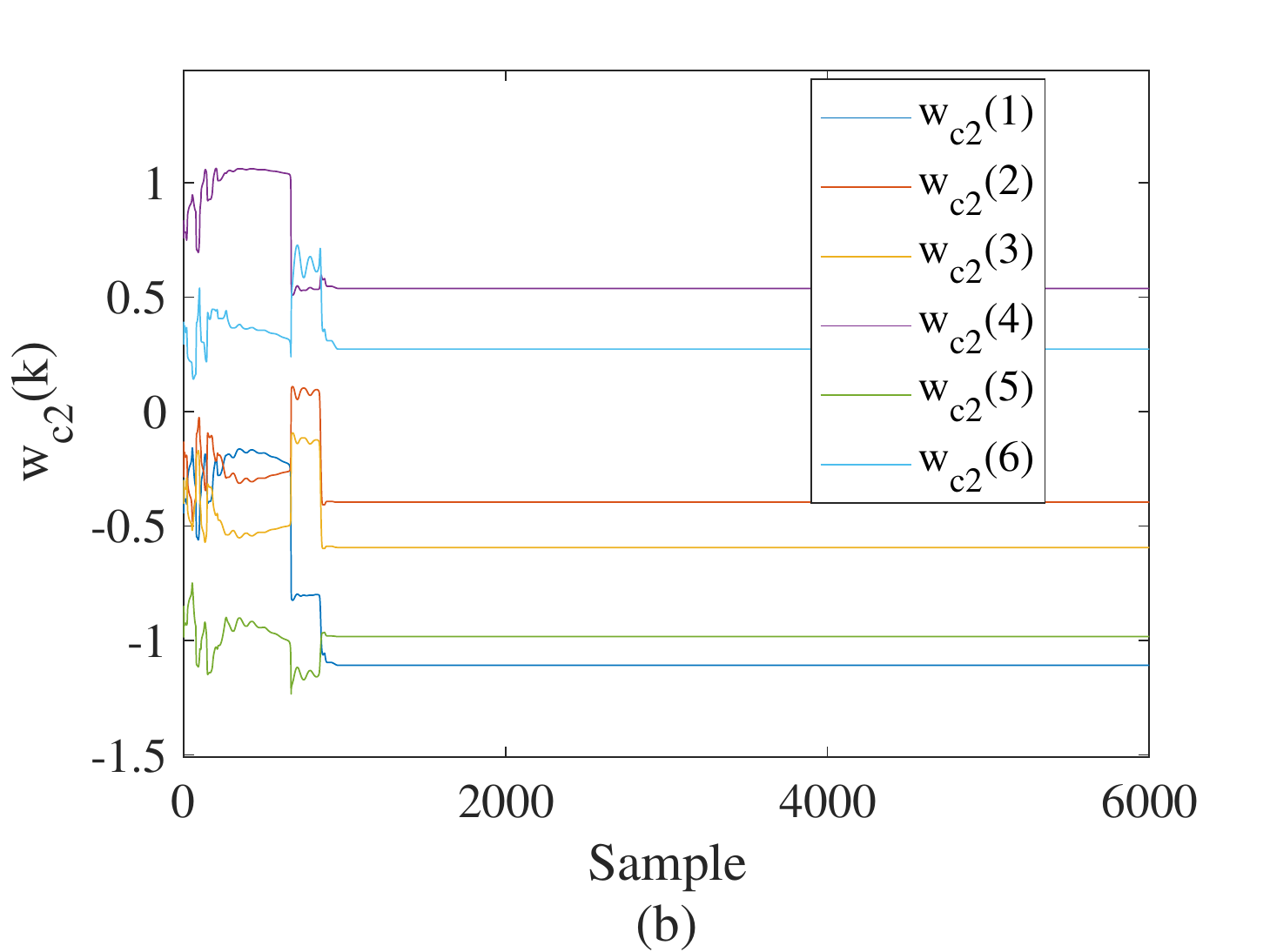}
\end{minipage}
\end{tabular}
\caption{The weights of action network with backstepping (a) and the weights of critic network with backstepping (b).}
\end{figure}

\subsection{Reproducibility of the proposed scheme}
While the previous evaluation has illustrated the effectiveness of the proposed tracking control design, we are now in a position to perform a comprehensive evaluation. To do so, nine different scenarios (Table \uppercase\expandafter{\romannumeral1}) are used to quantitatively evaluate the reproducibility of results using the proposed algorithm. In obtaining the results, a trial is successful if the MSE$_{3000^{+}}$ is less than that of feedback stabilizing control method in one trial. In Table \uppercase\expandafter{\romannumeral1}, ``C" denotes a constant 5, ``V" a constant 5 plus random Gaussian with mean = 0 and std = 0.50, a ``Pulse" disturbance is $\tau_{d}(t)=2$ appearing only at $t=40$,
and ``Gaus" represents Gaussian noise with mean = 0 and std = 8.25, respectively. The success rate over 50 trials is shown in Table I where the reproducibility of the proposed algorithm is verified.

\begin{figure}[ht]
	\centering
	\includegraphics[scale=0.5]{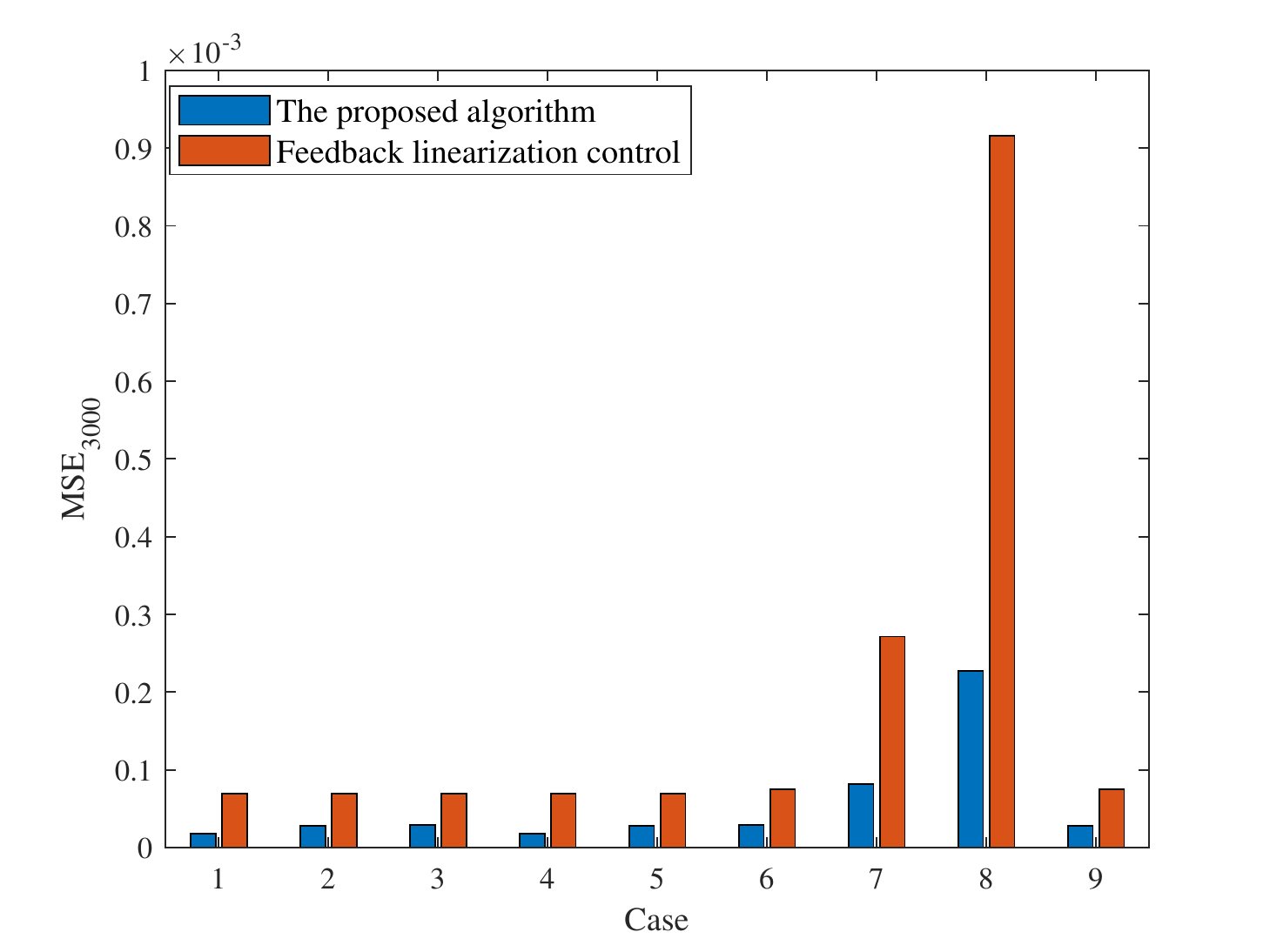}
	\caption{Comparison of average MSE$_{3000^{+}}$.}
	\label{fig:labe7}
\end{figure}

\subsection{Reproducibility with reset after failure}

To test if the proposed method can lead to 100\% success, we use a ``reset” mechanism. Previously we defined a trial as a simulation containing 6000 time samples starting from a given initial state and randomly initialized weights in ANN and CNN. Now we define an episode containing multiple trials until reaching success. Each trial in an episode has 6000 time samples. An episode starts from a given initial state and randomly initialized ANN and CNN weights. But if a trial ends with a failure, only the ANN and CNN weights are reset to random values for the next trial. The ``\# of reset" in Table Table \uppercase\expandafter{\romannumeral1} is the number of resets occurred in 50 episodes.  Note that, the ``success rate” column reports results from subsection B in the above. The average MSE$_{3000^{+}}$ of the last successful trial over the 50 episodes is shown in Fig. 4. For comparison, we also show the MSE$_{3000^{+}}$ from applying feedback stabilizing control. It can be seen that the proposed algorithm outperforms feedback stabilizing control.

Example 2. We now consider a two-link robot manipulator with the following motion equation
\begin{eqnarray}
M(q(t)) \ddot{q}(t) + V_{m}(q(t), \dot{q}(t)) \dot{q}(t) + F_d + F_s \dot{q}(t) + \tau_{d}(t) = \tau(t).
\end{eqnarray}
where $q(t)=[q_{1}(t) \quad q_{2}(t)]^T$, the inertia  matrix is given by
$$
M(q(t)) \triangleq\left[\begin{array}{cc}p_{1}+2 p_{3} s_{1}(t) & p_{2}+p_{3} s_{1}(t) \\ p_{2}+p_{3} s_{1}(t) & p_{2}\end{array}\right],$$
the centripetal-coriolis  matrix is given by
$$
V_{m}(q(t), \dot{q}(t)) \triangleq\left[
\begin{array}{cc}-p_{3} s_{2}(t) \dot{q}_{2}(t) & -p_{3} s_{2}(t)\left(\dot{q}_{1}(t)+\dot{q}_{2}(t)\right) \\ p_{2} s_{2}(t) \dot{q}_{1}(t) & 5\end{array}\right],$$
where $s_{1}=\cos \left(q_{2}\right)$, $s_{2}=\sin \left(q_{2}\right)$;
$p_{1}=3.473 \mathrm{~kg} \mathrm{~m}^{2}$, $p_{2}=0.196 \mathrm{~kg} \mathrm{~m}^{2}$, and $p_{3}=0.242 \mathrm{~kg} \mathrm{~m}^{2}$; $F_{d}=\mathrm{diag}[5.3, 6.1] \mathrm{N} ~ \mathrm{m} ~ \mathrm{s}$ and $F_{s}(\dot{q})=[8.45 \tanh (\dot{q}_{1}), ~2.35 \tanh (\dot{q}_{2})]^{T} \mathrm{~N} \mathrm{~m}$ are the models for the static and the dynamic friction, respectively.

The feedback gain parameters $c_{1}$ and $c_{2}$ in Eq. (64) are chosen as $c_{1}= 0.7$ and $c_{2}= -1$. The CNN and ANN in Fig.1 each has eight hidden nodes. The discount factor $\gamma$ in Eq. (16) is chosen as 0.95, the initial states are $q(t)=[1.8 \quad 1.5]^T$ and $\dot{q}(t)=[0 \quad 0]^T$ and the learning rate $l_{a}$, $l_{c}$ are 0.01. We use a sampling time period of $h=0.01s$ and the initialization of $M^{+}(k)$ is 
$[p_{1} ~ p_{2}; p_{2} ~ p_{2}]/h$.

A comparison of tracking performance shows that the MSE$_{3000^{+}}$ of the stabilizing control is $2.69\times10^{-2}$ for $q_{1}$ and $1.54\times10^{-2}$ for $q_{2}$ while the MSE$_{3000^{+}}$ of the dHDP-based is $0.19\times10^{-2}$ for $q_{1}$ and $0.21\times10^{-2}$ for $q_{2}$, respectively. This example again verifies the effectiveness of the proposed tracking control design.

% \begin{figure}
% 	\centering
% 	\includegraphics[width=250pt]{fig8.eps}\\
% 	\caption{{\color{blue}Tracking controlperformance comparison by feedback stabilizing control (a) with MSE$_{3000^{+}}=2.69\times10^{-2}$ and (c) with MSE$_{3000^{+}}=1.54\times10^{-2}$, and by dHDP-based control (b) with MSE$_{3000^{+}}=0.19\times10^{-2}$ and (d) with MSE$_{3000^{+}}=0.21\times10^{-2}$.}}
% 	\label{fig:feature extration}
% \end{figure}

\section{Conclusion}

This study aims at developing a mathematically suitable and practically useful, data-driven tracking control solution. Toward this goal, we introduce a new dHDP-based tracking control algorithm that takes advantage of the potential closed-loop system stability framework of the backstepping design for Euler-Lagrange systems. Such design approach also removes the dependence on a reference model for the desired tracking trajectory. Based on the proposed algorithm, we have shown stability of the overall dynamic system, weight convergence of the actor-critic neural networks, and (sub)optimality of the Bellman solution. Our simulations  show improved reproducibility and tracking error of the tracking control design. As the dHDP has been shown feasible to solve complex engineering application problems, it is expected that this algorithm also has the potential for applications of tracking control of nonlinear dynamic systems.

\bibliographystyle{IEEEtran}

\bibliography{TowardsReliableDesigns}

\begin{thebibliography}{10}
\providecommand{\url}[1]{#1}
\csname url@samestyle\endcsname
\providecommand{\newblock}{\relax}
\providecommand{\bibinfo}[2]{#2}
\providecommand{\BIBentrySTDinterwordspacing}{\spaceskip=0pt\relax}
\providecommand{\BIBentryALTinterwordstretchfactor}{4}
\providecommand{\BIBentryALTinterwordspacing}{\spaceskip=\fontdimen2\font plus
\BIBentryALTinterwordstretchfactor\fontdimen3\font minus
  \fontdimen4\font\relax}
\providecommand{\BIBforeignlanguage}[2]{{%
\expandafter\ifx\csname l@#1\endcsname\relax
\typeout{** WARNING: IEEEtran.bst: No hyphenation pattern has been}%
\typeout{** loaded for the language `#1'. Using the pattern for}%
\typeout{** the default language instead.}%
\else
\language=\csname l@#1\endcsname
\fi
#2}}
\providecommand{\BIBdecl}{\relax}
\BIBdecl

\bibitem{krstic1995nonlinear}
M.~Krstic, P.~V. Kokotovic, and I.~Kanellakopoulos, \emph{Nonlinear and
  adaptive control design}.\hskip 1em plus 0.5em minus 0.4em\relax John Wiley
  \& Sons, Inc., 1995.

\bibitem{khalil2002nonlinear}
H.~K. Khalil and J.~W. Grizzle, \emph{Nonlinear systems}.\hskip 1em plus 0.5em
  minus 0.4em\relax Prentice hall Upper Saddle River, NJ, 2002, vol.~3.

\bibitem{nijmeijer1990nonlinear}
H.~Nijmeijer and A.~Van~der Schaft, \emph{Nonlinear dynamical control
  systems}.\hskip 1em plus 0.5em minus 0.4em\relax Springer, 1990, vol. 175.

\bibitem{isidori2013nonlinear}
A.~Isidori, \emph{Nonlinear control systems}.\hskip 1em plus 0.5em minus
  0.4em\relax Springer Science \& Business Media, 2013.

\bibitem{sun2019novel}
K.~Sun, J.~Qiu, H.~R. Karimi, and H.~Gao, ``A novel finite-time control for
  nonstrict feedback saturated nonlinear systems with tracking error
  constraint,'' \emph{IEEE Transactions on Systems, Man, and Cybernetics:
  Systems}, 2019.

\bibitem{fan2018robust}
B.~Fan, Q.~Yang, X.~Tang, and Y.~Sun, ``Robust adp design for continuous-time
  nonlinear systems with output constraints,'' \emph{IEEE Transactions on
  Neural Networks and Learning Systems}, vol.~29, no.~6, pp. 2127--2138, 2018.

\bibitem{fu2020mrac}
H.~Fu, X.~Chen, W.~Wang, and M.~Wu, ``Mrac for unknown discrete-time nonlinear
  systems based on supervised neural dynamic programming,''
  \emph{Neurocomputing}, vol. 384, pp. 130--141, 2020.

\bibitem{radac2019data}
M.-B. Radac and R.-E. Precup, ``Data-driven model-free tracking reinforcement
  learning control with vrft-based adaptive actor-critic,'' \emph{Applied
  Sciences}, vol.~9, no.~9, p. 1807, 2019.

\bibitem{zhang2008novel}
H.~Zhang, Q.~Wei, and Y.~Luo, ``A novel infinite-time optimal tracking control
  scheme for a class of discrete-time nonlinear systems via the greedy hdp
  iteration algorithm,'' \emph{IEEE Transactions on Systems, Man, and
  Cybernetics, Part B (Cybernetics)}, vol.~38, no.~4, pp. 937--942, 2008.

\bibitem{yang2009direct}
L.~Yang, J.~Si, K.~S. Tsakalis, and A.~A. Rodriguez, ``Direct heuristic dynamic
  programming for nonlinear tracking control with filtered tracking error,''
  \emph{IEEE Transactions on Systems, Man, and Cybernetics, Part B
  (Cybernetics)}, vol.~39, no.~6, pp. 1617--1622, 2009.

\bibitem{zhang2011data}
H.~Zhang, L.~Cui, X.~Zhang, and Y.~Luo, ``Data-driven robust approximate
  optimal tracking control for unknown general nonlinear systems using adaptive
  dynamic programming method,'' \emph{IEEE Transactions on Neural Networks},
  vol.~22, no.~12, pp. 2226--2236, 2011.

\bibitem{wei2014adaptive}
Q.~Wei and D.~Liu, ``Adaptive dynamic programming for optimal tracking control
  of unknown nonlinear systems with application to coal gasification,''
  \emph{IEEE Transactions on Automation Science and Engineering}, vol.~11,
  no.~4, pp. 1020--1036, 2014.

\bibitem{modares2014optimal}
H.~Modares and F.~L. Lewis, ``Optimal tracking control of nonlinear
  partially-unknown constrained-input systems using integral reinforcement
  learning,'' \emph{Automatica}, vol.~50, no.~7, pp. 1780--1792, 2014.

\bibitem{kiumarsi2015actor}
B.~Kiumarsi and F.~L. Lewis, ``Actor--critic-based optimal tracking for
  partially unknown nonlinear discrete-time systems,'' \emph{IEEE Transactions
  on Neural Networks and Learning Systems}, vol.~26, no.~1, pp. 140--151, 2015.

\bibitem{kamalapurkar2015approximate}
R.~Kamalapurkar, H.~Dinh, S.~Bhasin, and W.~E. Dixon, ``Approximate optimal
  trajectory tracking for continuous-time nonlinear systems,''
  \emph{Automatica}, vol.~51, pp. 40--48, 2015.

\bibitem{modares2015h}
H.~Modares, F.~L. Lewis, and Z.-P. Jiang, ``H tracking control of completely
  unknown continuous-time systems via off-policy reinforcement learning,''
  \emph{IEEE Transactions on Neural Networks and Learning Systems}, vol.~26,
  no.~10, pp. 2550--2562, 2015.

\bibitem{luo2016model}
B.~Luo, D.~Liu, T.~Huang, and D.~Wang, ``Model-free optimal tracking control
  via critic-only q-learning,'' \emph{IEEE Transactions on Neural Networks and
  Learning Systems}, vol.~27, no.~10, pp. 2134--2144, 2016.

\bibitem{mu2017data}
C.~Mu, Z.~Ni, C.~Sun, and H.~He, ``Data-driven tracking control with adaptive
  dynamic programming for a class of continuous-time nonlinear systems,''
  \emph{IEEE Transactions on Cybernetics}, vol.~47, no.~6, pp. 1460--1470,
  2017.

\bibitem{gao2018learning}
W.~Gao and Z.-P. Jiang, ``Learning-based adaptive optimal tracking control of
  strict-feedback nonlinear systems,'' \emph{IEEE Transactions on Neural
  Networks and Learning Systems}, vol.~29, no.~6, pp. 2614--2624, 2018.

\bibitem{wang2018neural}
D.~Wang, D.~Liu, Y.~Zhang, and H.~Li, ``Neural network robust tracking control
  with adaptive critic framework for uncertain nonlinear systems,''
  \emph{Neural Networks}, vol.~97, pp. 11--18, 2018.

\bibitem{zhao2019event}
B.~Zhao and D.~Liu, ``Event-triggered decentralized tracking control of modular
  reconfigurable robots through adaptive dynamic programming,'' \emph{IEEE
  Transactions on Industrial Electronics}, vol.~67, no.~4, pp. 3054--3064,
  2019.

\bibitem{mu2019learning}
C.~Mu and Y.~Zhang, ``Learning-based robust tracking control of quadrotor with
  time-varying and coupling uncertainties,'' \emph{IEEE Transactions on Neural
  Networks and Learning Systems}, vol.~31, no.~1, pp. 259--273, 2019.

\bibitem{dong2020optimal}
H.~Dong, X.~Zhao, and B.~Luo, ``Optimal tracking control for uncertain
  nonlinear systems with prescribed performance via critic-only adp,''
  \emph{IEEE Transactions on Systems, Man, and Cybernetics: Systems}, 2020.

\bibitem{ni2013adaptive}
Z.~Ni, H.~He, and J.~Wen, ``Adaptive learning in tracking control based on the
  dual critic network design,'' \emph{IEEE Transactions on Neural Networks and
  Learning Systems}, vol.~24, no.~6, pp. 913--928, 2013.

\bibitem{si2001online}
J.~Si and Y.-T. Wang, ``Online learning control by association and
  reinforcement,'' \emph{IEEE Transactions on Neural networks}, vol.~12, no.~2,
  pp. 264--276, 2001.

\bibitem{yang2009performance}
L.~Yang, J.~Si, K.~S. Tsakalis, and A.~A. Rodriguez, ``Performance evaluation
  of direct heuristic dynamic programming using control-theoretic measures,''
  \emph{Journal of Intelligent and Robotic Systems}, vol.~55, no. 2-3, pp.
  177--201, 2009.

\bibitem{Nhan2018Model}
N.~T. Nguyen, \emph{Model Reference Adaptive Control: A Primer}.\hskip 1em plus
  0.5em minus 0.4em\relax Springer, 2018.

\bibitem{da2019choice}
G.~R.~G. da~Silva, A.~S. Bazanella, and L.~Campestrini, ``On the choice of an
  appropriate reference model for control of multivariable plants,'' \emph{IEEE
  Transactions on Control Systems Technology}, vol.~27, no.~5, pp. 1937--1949,
  2019.

\bibitem{zargarzadeh2015optimal}
H.~Zargarzadeh, T.~Dierks, and S.~Jagannathan, ``Optimal control of nonlinear
  continuous-time systems in strict-feedback form,'' \emph{IEEE Transactions on
  Neural Networks and Learning Systems}, vol.~26, no.~10, pp. 2535--2549, 2015.

\bibitem{wang2016backstepping}
Z.~Wang, X.~Liu, K.~Liu, S.~Li, and H.~Wang, ``Backstepping-based lyapunov
  function construction using approximate dynamic programming and sum of square
  techniques,'' \emph{IEEE Transactions on Cybernetics}, vol.~47, no.~10, pp.
  3393--3403, 2016.

\bibitem{miranda2017multivalued}
F.~A. Miranda-Villatoro, B.~Brogliato, and F.~Castanos, ``Multivalued robust
  tracking control of lagrange systems: Continuous and discrete-time
  algorithms,'' \emph{IEEE Transactions on Automatic Control}, vol.~62, no.~9,
  pp. 4436--4450, 2017.

\bibitem{si2004handbook}
J.~Si, A.~G. Barto, W.~B. Powell, and D.~Wunsch, \emph{Handbook of learning and
  approximate dynamic programming}.\hskip 1em plus 0.5em minus 0.4em\relax John
  Wiley \& Sons, 2004, vol.~2.

\bibitem{enns2002apache}
R.~Enns and J.~Si, ``Apache helicopter stabilization using neural dynamic
  programming,'' \emph{Journal of Guidance, Control, and Dynamics}, vol.~25,
  no.~1, pp. 19--25, 2002.

\bibitem{enns2003helicopterj}
------, ``Helicopter flight-control reconfiguration for main rotor actuator
  failures,'' \emph{Journal of Guidance, Control, and Dynamics}, vol.~26,
  no.~4, pp. 572--584, 2003.

\bibitem{enns2003helicopter}
------, ``Helicopter trimming and tracking control using direct neural dynamic
  programming,'' \emph{IEEE Transactions on Neural networks}, vol.~14, no.~4,
  pp. 929--939, 2003.

\bibitem{lu2008direct}
C.~Lu, J.~Si, and X.~Xie, ``Direct heuristic dynamic programming for damping
  oscillations in a large power system,'' \emph{IEEE Transactions on Systems,
  Man, and Cybernetics, Part B (Cybernetics)}, vol.~38, no.~4, pp. 1008--1013,
  2008.

\bibitem{wen2017new}
Y.~Wen, J.~Si, X.~Gao, S.~Huang, and H.~Huang, ``A new powered lower limb
  prosthesis control framework based on adaptive dynamic programming.''
  \emph{IEEE Transactions on Neural Networks and Learning Systems}, vol.~28,
  no.~9, pp. 2215--2220, 2017.

\bibitem{wen2019online}
Y.~Wen, J.~Si, A.~Brandt, X.~Gao, and H.~Huang, ``Online reinforcement learning
  control for the personalization of a robotic knee prosthesis,'' \emph{IEEE
  Transactions on Cybernetics}, 2019.

\bibitem{zhang2019adaptive}
Y.~Zhang, S.~Li, K.~J. Nolan, and D.~Zanotto, ``Adaptive assist-as-needed
  control based on actor-critic reinforcement learning.'' in \emph{The IEEE/RSJ
  International Conference on Intelligent Robots and Systems (IROS)}, 2019, pp.
  4066--4071.

\bibitem{liu2012boundedness}
F.~Liu, J.~Sun, J.~Si, W.~Guo, and S.~Mei, ``A boundedness result for the
  direct heuristic dynamic programming,'' \emph{Neural Networks}, vol.~32, pp.
  229--235, 2012.

\bibitem{sokolov2015complete}
Y.~Sokolov, R.~Kozma, L.~D. Werbos, and P.~J. Werbos, ``Complete stability
  analysis of a heuristic approximate dynamic programming control design,''
  \emph{Automatica}, vol.~59, pp. 9--18, 2015.

\bibitem{michel2008stability}
A.~N. Michel, L.~Hou, and D.~Liu, \emph{Stability of dynamical systems}.\hskip
  1em plus 0.5em minus 0.4em\relax Springer, 2008.

\bibitem{sarangapani2018neural}
J.~Sarangapani, \emph{Neural network control of nonlinear discrete-time
  systems}.\hskip 1em plus 0.5em minus 0.4em\relax CRC press, 2018.

\end{thebibliography}

\end{document}